\documentclass[pre,superscriptaddress,twocolumn]{revtex4-1}
\usepackage[utf8]{inputenc}
\usepackage{amsmath,amssymb,amsfonts,graphicx}
\usepackage[colorlinks=true,allcolors=blue]{hyperref}
\usepackage{dsfont}
\usepackage[english]{babel}

\usepackage[normalem]{ulem}
\usepackage[dvipsnames]{xcolor}

\begin{document}

\title{Autocorrelation properties of temporal networks governed by dynamic node variables}

\author{Harrison Hartle}
\email{hthartle1@gmail.com}
\affiliation{Santa Fe Institute, Santa Fe, NM 87501, USA}

\author{Naoki Masuda}
\email{naokimas@gmail.com}
\affiliation{Department of Mathematics, State University of New York at Buffalo, NY 14260-2900, USA}
\affiliation{Institute for Artificial Intelligence and Data Science, University at Buffalo, State University of New York at Buffalo, Buffalo, NY 14260-5030, USA}
\affiliation{Center for Computational Social Science, Kobe University, Kobe, 657-8501, Japan}

\date{\today}

\begin{abstract}
We study synthetic temporal networks whose evolution is determined by stochastically evolving node variables -- synthetic analogues of, e.g.,  temporal proximity networks of mobile agents. We quantify the long-timescale correlations of these evolving networks by an autocorrelative measure of edge persistence. Several distinct patterns of autocorrelation arise, including power-law decay and exponential decay, depending on the choice of node-variable dynamics and connection probability function. Our methods are also applicable in wider contexts; our temporal network models are tractable mathematically and in simulation, and our long-term memory quantification is analytically tractable and straightforwardly computable from temporal network data. 
\end{abstract}

\maketitle

\section{Introduction}
\label{sec:intro}

In dynamic networks, edges can evolve, and so too can node variables such as spatial positions, group memberships, and popularity levels. These node variables in some instances encode the connection compatibility among nodes, as described by latent-variable network models \cite{bianconi2001bose,caldarelli2002scale,boguna2003class,chung2003spectra,masuda2004analysis}. This observation motivates a class of temporal network models derived as dynamic analogues to such latent-variable models, with examples including proximity networks (random geometric graphs, RGGs) \cite{penrose2003random}, popularity-based attachment rules (e.g., hypersoft configuration models) \cite{caldarelli2002scale,van2018sparse}, group-affinity based attachment (e.g., stochastic block models, SBMs)  and models in which multiple node variables influence connectivity (e.g., in hyperbolic random graphs, the $\mathbb{S}^1$ model, and degree-corrected SBMs) \cite{serrano2008self,krioukov2010hyperbolic,karrer2011stochastic}, generalized by the notion of hidden-variable model class \cite{boguna2003class}. Numerous temporal extensions of static hidden-variable network models have been derived; examples include SBMs with dynamic community assignments \cite{barucca2018disentangling}, RGGs with dynamic coordinates \cite{diaz2007dynamic}, hyperbolic random graphs with dynamic popularity and similarity-encoding variables \cite{papaefthymiou2024fundamental}, and other related models \cite{friel2016interlocking,mazzarisi2020dynamic,ward2013gravity,sewell2015latent}. 

In this category of models -- temporal networks influenced by dynamic node variables -- many results have been obtained, such as the characterization of percolation properties \cite{peres2013mobile}, identification of change points in group-affiliation configurations \cite{peixoto2023modelling}, quantification of deviations from the corresponding static models \cite{hartle2021dynamic}, and inference of node-variable trajectories \cite{ghasemian2016detectability}. These types of temporal network models can describe real-world systems including (i) temporal proximity graphs among mobile agents such as humans \cite{panisson2013fingerprinting,scholz2014predictability}, vehicles \cite{hartenstein2008tutorial}, nonhuman organisms \cite{morales2010building}, and autonomous robots \cite{fink2013robust}, (ii) temporal economic/trade networks (influenced by dynamic node variables such as supply, demand, geography, and firm strategies) \cite{de2011world,volberda2003co,kennedy2023building}, and (iii) temporal ecological networks on evolutionary timescales (influenced by genetic profiles and population sizes of species, which can be abstracted as dynamic node variables) \cite{carroll2007evolution,held2014adaptive,hendry2017eco}, among others. These real systems motivate further theoretical and empirical investigation of how {\it dynamic} node variables influence {\it dynamic} networks.

In the present study, we propose several models in which dynamic node variables give rise to temporal networks, for the purpose of examining how the choice of stochastic process governing node variables impacts the {\it dynamics} of the network (as opposed to the structure of individual snapshots). We therefore (i) select a collection of temporal network models with simple {\it static} properties, and with a deterministic relationship between node variables and network structure at any given time, and (ii) select an appropriate temporal network quantifier, namely, a measure of long-timescale network-structural memory: the autocorrelation of pairwise connectivity. In the remainder of this section, we describe and justify our choices of dynamic network models (Sec.~\ref{ssec:modeling_choices}) and of temporal network properties of interest (Sec.~\ref{ssec:property_choices}). We then explain how our work relates to past studies (Sec.~\ref{ssec:past_work}) and finally state the organization of the rest of this article (Sec.~\ref{ssec:article_organization}).

\subsection{Choice of temporal network models}
\label{ssec:modeling_choices}

We consider random temporal networks in which node variables evolve by independent stochastic processes with random initial conditions. We assume that the network structure, or more specifically, the presence or absence of an edge between two nodes, is a deterministic function of the time-varying node variables of those two nodes (e.g., by proximity thresholds). These assumptions are for analytical and conceptual simplicity. As a result, we have simple network {\it structure} at any time (e.g., being composed of a collection of cliques, or being statistically equivalent to a static RGG). Yet a nontrivial network {\it dynamics} arises through the dependency on stochastic node-variable dynamics, which we explore.

\subsection{Choice of temporal network descriptor}
\label{ssec:property_choices}

Temporal networks are quantifiable by many measures \cite{holme2012temporal, masuda2020guide}. Rather than considering measures applicable to individual timestep, we examine instead how dynamical snapshots relate to one another. In particular, we examine the long-timescale network-structural memory as quantified by persistence $R(s)$, defined as the probability that an edge exists at time $t+s$ given that it exists at time $t$, regardless of whether it existed at intermediate times, and averaged across all node-pairs and initial times $t$. By definition, it holds that $R(0)=1$. Under stationarity assumptions, we obtain $\lim_{s\to\infty} R(s)=\langle \rho\rangle$, where $\langle\rho\rangle$ is the stationary probability with which any given edge exists (i.e., the edge density). The pairwise nature of $R(s)$ renders its computation analytically tractable in the models considered herein.

The persistence $R(s)$ is proportional to the autocorrelation of the connectivity time series (namely, the binary sequence indicating whether a pair is connected at each time $t$). Autocorrelative measures have been applied in diverse contexts \cite{bernasconi1987low,jones2012autocorrelation,sokal1978spatial,porte2014autocorrelation,agterberg1970autocorrelation,wise1955autocorrelation} and hence are a natural choice as temporal network descriptor, as have been considered in multiple past studies \cite{clauset2012persistence, williams2019effects, lacasa2022correlations, jo2024temporal}. Additionally, quantities related to $R(1)$ have been used as temporal network descriptors \cite{Nicosia2013,buttner2016adaption}.

\subsection{Further relations to past work}
\label{ssec:past_work}

Our work also involves other elements present in past studies, beyond those mentioned in Secs.~\ref{ssec:modeling_choices} and \ref{ssec:property_choices}. We briefly summarize these works here, making further mention of them in the discussion (Sec.~\ref{sec:discussion}).

Dynamic versions of random geometric graphs have been studied in a variety of contexts \cite{navidi2004stationary,sinclair2010mobile,fujiwara2011synchronization}, including the case of coordinates governed by L\'evy flights \cite{rhee2011levy, birand2011dynamic}. The latter modeling studies considered two-dimensional RGGs with nodes undergoing hops at irregular times, with both the hop-size distribution and waiting-time distribution being heavy-tailed, with hops lasting a nonzero duration, and with reflecting boundary conditions. 

Mobile geometric graphs, studied in the mathematics community \cite{sinclair2010mobile}, are dynamic proximity networks among infinitely many nodes sprinkled into $\mathbb{R}^D$ with uniform density, undergoing i.i.d.\,Brownian motion. In these past studies, the properties of interest include (i) percolation phenomena such as the existence of an  infinite component and whether that component exists for all time; (ii) properties of nodes in relation to the graph, such as the duration before isolated nodes become a part of the infinite component \cite{peres2013mobile}. These models arose from mathematical considerations of point processes -- namely, the ``dynamic Boolean model'' \cite{van1997dynamic,konstantopoulos2014mobile}, which considers the overlap properties among spheres at random locations with different radii, undergoing motion in the space.

In the study of vehicular ad hoc networks \cite{hartenstein2008tutorial}, and more broadly the field of mobile ad hoc networks \cite{ramanathan2002brief}, many models resembling those considered in the present paper have been studied \cite{zhang2014connectivity,shang2009exponential}. Numerous studies introduce dynamical processes atop such networks of mobile agents, including susceptible-infectious-recovered (SIR) -like dynamics \cite{grimmett2022brownian}, flooding processes \cite{clementi2015parsimonious}, gossip \cite{zhang2013gossip}, opinion models \cite{ferri2023three}, routing performance \cite{mauve2001survey,ye2003framework}, and others \cite{gu2018ability, cheliotis2020simple}.

\subsection{Article organization}
\label{ssec:article_organization}

In Sec.~\ref{sec:models}, we present the stochastic dynamic network models studied in the present study. In Sec.~\ref{sec:results}, we present our results on persistence $R(s)$ in each setting. In Sec.~\ref{sec:discussion}, we further discuss this work's context and follow-up prospects. Derivations of the analytical results appear in Appendices~\ref{app:brownian_Rs_calculation},~\ref{app:eta_Rs_alpha},~\ref{app:P_I_given_x_eq_x},~\ref{app:metapopulation_graph},~\ref{app:ANDmodel}, and \ref{app:recursion_solution}.

    \begin{figure}[h]
        \centering
        \includegraphics[scale=0.22,trim=160 40 30 150,clip=true]{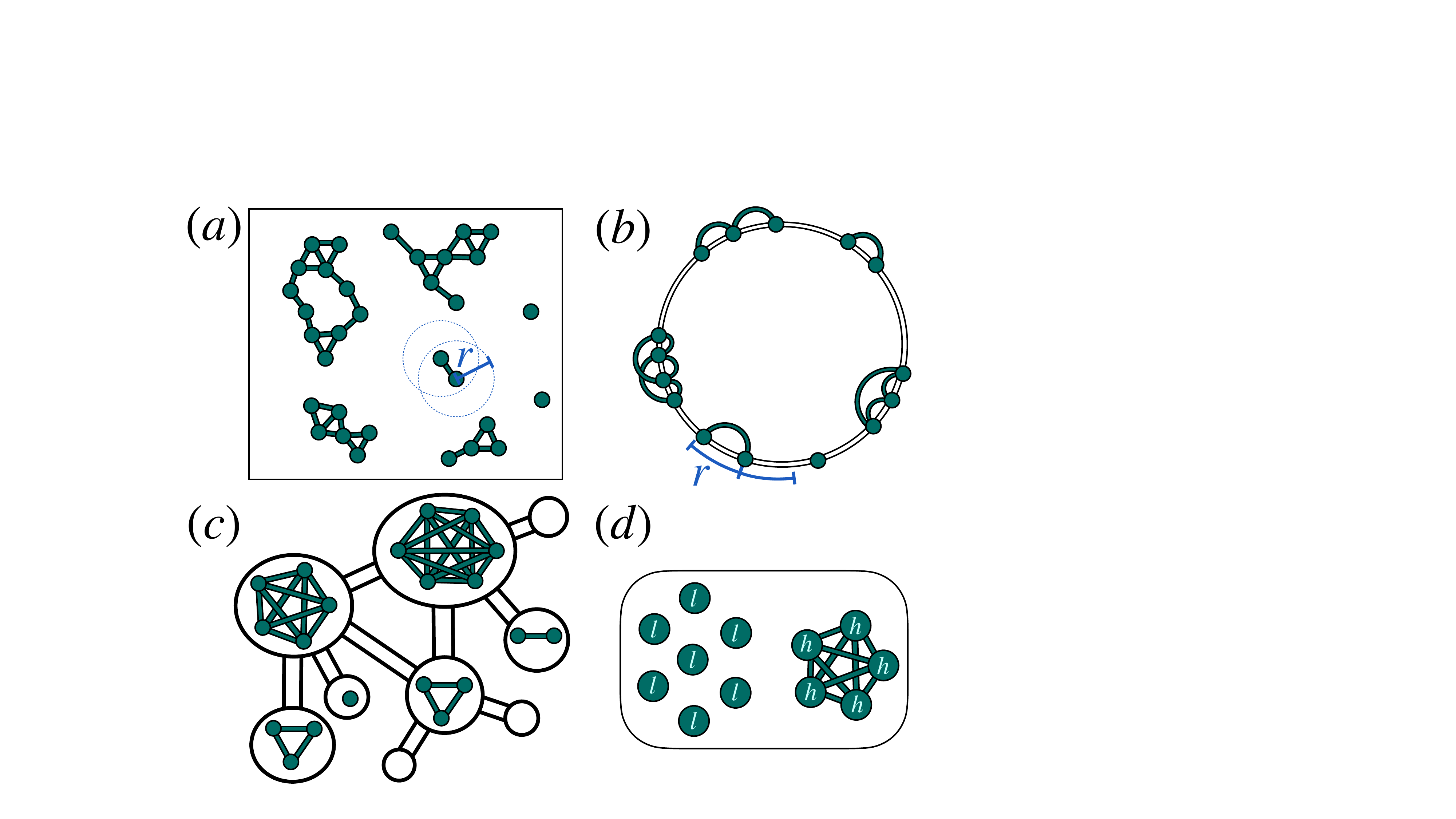}
        \caption{Network structure arising deterministically from node variables. (a) Two-dimensional random geometric graph; all node pairs whose distance is less than a threshold $r$ are adjacent to each other. (b) One-dimensional random geometric graph on the circle under the same threshold connection rule. (c) In the dynamic metapopulation proximity network, all individuals (i.e., nodes) occupying the same site in the metapopulation graph are adjacent to one another, and no other edges are present between individuals. (d) In the AND model, all nodes are either in state $\ell$ (``low'') or state $h$ (``high''), and all $h$ nodes are pairwise adjacent to form a clique, whereas all other node pairs are not adjacent. In (a) and (b), the node variables are continuous. In (c) and (d), the node variables are discrete.}
        \label{fig:structure_from_node_variables}
    \end{figure}

\section{Node-variable models for generating dynamic networks}
\label{sec:models}

In this section, we introduce dynamic network models. In all cases, we consider discrete time $t\in\{1, \ldots,T\}$, and time-dependent node variables denoted by $X^{(t)}=(X_i^{(t)})_{i=1}^n$, where $n$ is the number of nodes, and $X^{(t)}_i\in\mathcal{X}$ for a set $\mathcal{X}$. We denote the time-dependent adjacency matrix of the network by $A^{(t)}=(A_{ij}^{(t)})_{1\le i<j\le n}$, with $A_{ij}^{(t)}=1$ if the $i$th and $j$th nodes are adjacent at time $t$, and $A_{ij}^{(t)}=0$ otherwise. We assume that the network is undirected and unweighted. We also assume that $G^{(t)}$ is obtainable from $X^{(t)}$ through a deterministic mapping. We define three dynamic network models, one in each of the subsequent sections.

\subsection{Dynamic random geometric graphs}
\label{ssec:dynamic_RGGs}

In this section, we describe a dynamic variant of random geometric graphs (RGGs). Static RGGs have coordinates of the nodes sprinkled randomly into a space, $\mathcal{X}$ \cite{dall2002random,penrose2003random}. In the dynamic case, coordinates are initially randomly sampled and undergo a stochastic process in $\mathcal{X}$. As in the case of static RGGs, the dynamic RGGs are assumed to have edges defined by (instantaneous) proximity among coordinates; see Fig.~\ref{fig:structure_from_node_variables}a and \ref{fig:structure_from_node_variables}b. Precisely, we define edges by 
\begin{equation}
\begin{aligned}
\label{eq:RGG_connection_prob}
    A_{ij}^{(t)}=\mathds{1}\left\{d(X_i^{(t)},X_j^{(t)})\le r\right\},
\end{aligned}
\end{equation}
where $1\le i<j\le n$, $1\le t\le T$, $d$ is a distance metric among coordinates, and $\mathds{1}\{\mathcal{E}\}$ is the indicator function, equal to one if event $\mathcal{E}$ holds, and equal to zero otherwise.

For simplicity, we consider a one-dimensional circular space of circumference $n$ (see Fig.~\ref{fig:structure_from_node_variables}b), parameterized by the open interval $[0,n)$. Each initial coordinate $X_i^{(1)}\in[0,n]$ is sampled independently of all others from the uniform density $\mathrm{Unif}[0,n]$, where $\mathrm{Unif}$ represents the uniform density. Because $\mathrm{Unif}[0,n]$ is the stationary density of $X_i^{(t)}$ in each of the three stochastic dynamics of the node described in the following text, we obtain $X_i^{(t)}\sim \mathrm{Unif}[0,n]$ for any $i \in \{1, \ldots, n \}$ and $t \in \{ 2, 3, \ldots \}$ as well. Therefore, the temporal network $G^{(t)}$ is also in stationarity. The average degree in a given snapshot is denoted by $\bar{k}(G^{(t)})=\frac{1}{n}\sum_{i=1}^n\sum_{j\ne i}A_{ij}^{(t)}$. One can calculate the expectation of $\bar{k}(G^{(t)})$ by noting that, for any particular $i$, the variables $\{A_{ij}^{(t)}\}_{j\ne i}$ are independent Bernoulli variables of mean $\frac{2r}{n}$, which is the fractional volume of the circular space occupied by the connection radius in both directions. By stationarity, we have $\langle \bar{k}(G^{(t)})\rangle=\frac{2r}{n}(n-1)=2r(1-o(1))$ at any $t$.

We assume a fixed $r$ to maintain constant average degree as $n\rightarrow\infty$. At finite $n$, the distance metric incorporating periodic boundary conditions is
\begin{equation}\begin{aligned}\label{eq:dist_S1}
d(x,x')&=\frac{n}{2}-\left\vert\frac{n}{2}-|x-x'|\right\vert.
\end{aligned}\end{equation}
For $n\rightarrow\infty$, Eq.~\eqref{eq:dist_S1} becomes the ordinary Euclidean distance $d(x,x')=|x-x'|$ at any given $x$ and $x'$.

We use the following three stochastic processes governing coordinates of each node in the dynamic RGG. We provide further details in Sec.~\ref{ssec:computing_Rs_rgg} and in appendices referenced therein.
\begin{itemize}
\item {\it Teleportation dynamics}: coorindates remain fixed ($X_{i}^{(t)}=X_i^{(t-1)}$) except upon occasionally teleporting to a uniformly random location (i.e., by resampling $X_i^{(t)}\sim \mathrm{Unif}[0,n]$). The teleportation events arise independently with a given probability per unit time. See Sec.~\ref{sssec:tele}.
\item {\it Brownian motion}: coordinates undergo a random walk with Gaussian-distributed increments, or the Brownian motion. See Sec.~\ref{sssec:brownian}.
\item {\it L\'evy flights}: coordinates undergo an {\it $\alpha$-stable L\'evy flight}, whereby increments are drawn from an $\alpha$-stable distribution. L\'evy flights generalize Brownian motion, which corresponds to $\alpha=2$. See Sec.~\ref{sssec:levy}.
\end{itemize}

 The discrete-time processes we consider are equivalent to subsampled continuous-time processes at a set of continuous time values $\tilde{t}\in\{\tau i\}_{i=0}^{T-1}\subset \mathbb{R}_+$ for some time-resolution parameter $\tau>0$; those continuous time values are mapped to integer times $t\in\{1, \ldots,T\}$ via $\tilde{t}(t)=(t-1)\tau$. We rely on this continuous-time expression in many calculations in the following text. We write $X_i(\tilde{t})$ to denote dependence on continuous time, in contrast to $X_i^{(t)}$, which is defined to be in discrete time. All the three dynamical processes have a uniform distribution of coordinates in equilibrium, are memoryless, stationary, and are left-right symmetric (i.e., the displacement distribution is an even function).
 
 \subsection{Dynamic metapopulation proximity network}
 \label{ssec:metapop}

The next dynamic network model we consider is a dynamic metapopulation proximity network among mobile agents traversing the sites of a static metapopulation graph. Metapopulation graphs are commonly used for modeling disease and ecological dynamics \cite{hethcote1978immunization,may1984spatial,hanski1998metapopulation,colizza2007reaction}.

In this model, $n$ agents independently traverse an undirected metapopulation graph composed of $N$ sites (i.e., nodes of the metapopulation graph), following the simple random walk \cite{masuda2017random}. The node variable $X_i^{(t)} \in \{1, \ldots, N \}$ encodes the site on which the $i$th agent resides at time $t$.

Metapopulation graph models often assume that the agents can interact with each other only if they reside at the same site. If all agents on a common site interact with one another, the corresponding dynamic network is determined by
\begin{equation}
A_{ij}^{(t)}=\mathds{1}\left\{X_i^{(t)}=X_j^{(t)}\right\}.
\end{equation}
This model can be viewed as a dynamic RGG with connection radius $r=0$ on a metapopulation-structured space. At any time $t$, the network is composed of a collection of cliques, each composed of all agents visiting a common site (see Fig.~\ref{fig:structure_from_node_variables}c).

At each time step, each agent independently moves with probability $q$ to a neighboring site that is selected uniformly at random; the agent does not move with probability $1-q$. The initial location of the agent, $X_i^{(1)}$, is assigned i.i.d. at random from distribution $p_I=k_I/2M$, where $k_I$ denotes the degree of site $I$ of the metapopulation graph, and $M$ is the number of edges in the metapopulation graph. The probability $p_I$ is the stationary distribution of the simple random walk on the metapopulation graph.

\subsection{Discrete AND model}
\label{ssec:dynamic_and}

In this last model, each node has a binary-valued dynamic state $X_i^{(t)}\in\{ \ell, h\}$, evolving indepdendently and identically according to a Markov chain with stationary probabilities $u \equiv \mathbb{P}(X_i=h)$ and $1-u=\mathbb{P}(X_i=\ell)$. We define the adjacency matrix at any time $t$ by
\begin{equation}
    A_{ij}^{(t)}=\mathds{1}\left\{X_i^{(t)}=X_j^{(t)}=h\right\}.
\end{equation}
In other words, the two nodes are adjacent if and only if they are both in $h$, which stands for the high activity state. At any $t$, the resulting temporal network is composed of a clique among all the nodes in state $h$, with all the nodes in state $\ell$ being isolated (see Fig.~\ref{fig:structure_from_node_variables}d). At each time $t$, this network is a special case of the threshold networks \cite{masuda2005geographical,hagberg2006designing,caldarelli2002scale}. The edge density is given by $\langle \rho\rangle=\mathbb{P}(X_i^{(t)}=X_j^{(t)}=h)=u^2$. This model is not strictly geometric because, if $X_i(t)=X_j(t)= \ell$, then $A_{ij}^{(t)}=0$, indicating that proximity does not imply an edge. 

We sample the initial state of each node i.i.d. from the Bernoulli distribution with stationary probability $u$ of the $h$ state. (Therefore, the node's state is initially $\ell$ with probability $1-u$.) Thereafter, in each timestep, a node in state $\ell$ switches to state $h$ with probability $P_{\ell h}$ and remains in state $\ell$ with probability
$1 - P_{\ell h}$, independently of the other nodes. Similarly, a node in state $h$ switches to state $\ell$ with probability $P_{h\ell}$ and remains in state $h$ with probability $1 - P_{h\ell}$.
To respect the stationary probability, we impose
\begin{equation}
\frac{P_{\ell h}}{P_{h\ell}+P_{\ell h}}=u.
\end{equation}

\section{Results}
\label{sec:results}

Each model under consideration generates random temporal networks $G=(A^{(t)})_{t=1}^T$. In each model, we calculate the persistence, $R(s)$, defined as
\begin{equation}
\label{eq:Rs_definition}
R(s)=\mathbb{P}\left(\left. A_{ij}^{(t+s)}=1\right\vert A_{ij}^{(t)}=1\right),
\end{equation}
where $\mathbb{P}$ denotes the probability.
For the models under consideration, $R(s)$ is the same for all $1\le t\le T-s$ and $1\le i<j\le n$ due to the models' stationarity and statistical symmetry of node-pairs. The persistence is related to the autocorrelation
\begin{equation}
C(s)=\left\langle A_{ij}^{(t)} A_{ij}^{(t+s)}\right\rangle
\end{equation}
by
\begin{equation}
C(s)=\langle\rho\rangle R(s),
\end{equation}
where we recall that $\langle\rho\rangle$ is the edge density in stationarity. We study $R(s)$ rather than $C(s)$ for its more straightfoward interpretability in terms of network structure. Note that $R(0)=1$ and $R(\infty)=\langle \rho\rangle$. We consider asymptotics of $R(s)$ for $s\gg 1$ in the $n\rightarrow\infty$ regime. In sparse networks, it holds true that $\lim_{n\to\infty}\langle\rho\rangle =  0$ and consequently $R(s)\rightarrow 0$ for $s, n \rightarrow\infty$.

Our main results are summarized as follows. The first three results pertain to the three choices of coordinate dynamics for dynamics RGGs. In the case of coordinates undergoing teleportation dynamics, we obtain
\begin{equation}
\label{eq:Rs_tele_lam}
R(s) \propto e^{-\lambda s}, \ \lambda>0,
\end{equation}
where $\propto$ represents proportionality to leading order in the limit of $s\to\infty$. In the case of coordinates undergoing Brownian motion, we obtain
\begin{equation}
R(s) \propto s^{-\frac{1}{2}}.
\end{equation}
\label{eq:Rs_levy}
In the case of coordinates undergoing L\'evy flights with $\alpha\in(1,2)$, we obtain
\begin{equation}
R(s) \propto s^{-\frac{1}{\alpha}}, \ \alpha\in (1,2).
\end{equation}
For dynamic metapopulation proximity networks, we obtain for some coefficients $\{\mu_i\}$ and positive values $\{\lambda_i\}$,
\begin{equation}
\label{eq:Rs_meta_exp}
R(s) \propto \sum_{i}\mu_i e^{-\lambda_i s}.
\end{equation}
For the dynamic AND model, we obtain
\begin{equation}
R(s) \propto \xi^s+C\xi^{2s}, \ \ C>0,|\xi|<1.
\end{equation}
\label{theo:theorem1}

In what follows, we describe the derivation of $R(s)$ for each dynamic network model, referring to appendices for detailed proofs. 

\subsection{Computing $R(s)$ in dynamic RGGs}
\label{ssec:computing_Rs_rgg}

In dynamic RGGs (Sec.~\ref{ssec:dynamic_RGGs}), due to the deterministic proximity-based connection criterion (Eq.~\eqref{eq:RGG_connection_prob}), the persistence can be recast directly in terms of pairwise distances as follows:
\begin{equation}
\label{eq:Rs_formula_tele}
R(s)=\mathbb{P}\left(d_{ij}^{(t+s)}\le r\left\vert d_{ij}^{(t)} \le r \right.\right),
\end{equation}
where $d_{ij}^{(t)}=d( X_i^{(t)},X_j^{(t)} )$. In the one-dimensional case that we consider, Eq.~\eqref{eq:dist_S1} defines the distance metric $d$. In what follows, we describe how Eq.~\eqref{eq:Rs_formula_tele} is evaluated in the cases of teleportation dynamics (Sec.~\ref{sssec:tele}), Brownian motion (Sec.~\ref{sssec:brownian}), and ($\alpha$-stable) L\'evy flights with $\alpha\in(1,2)$ (Sec.~\ref{sssec:levy}).

\subsubsection{Random teleportation dynamics}
\label{sssec:tele}

In this section, we describe how $R(s)$ is computed in the case of random teleportation dynamics. Suppose a node pair is adjacent at time $t$. The node pair remains adjacent at time $t+s$ if neither node has teleported since time $t$. If either node has ever jumped, the two nodes are thereafter adjacent with probability $\langle\rho\rangle=2r/(n-1)$. Therefore, we obtain
\begin{equation}
\label{eq:Rs_tele}
R(s) = (1-\eta)^{2s}+\left[1-(1-\eta)^{2s}\right]\frac{2r}{n-1},
\end{equation}
where $\eta\in[0,1]$ is the probability of teleporting each time step. In the sparse regime, $R(s)\propto(1-\eta)^{2s}=e^{-\lambda s}$ with $\lambda=2\log(1-\eta)$.

\subsubsection{Brownian motion}
\label{sssec:brownian}

In this section, we describe the calculation of $R(s)$ in the case of Brownian motion. The relative position of the two nodes on the circle, 
\begin{equation}
Y(t) \equiv X_i^{(t)}-X_j^{(t)},
\end{equation}
is uniformly distributed in $[-n,n]$. Conditioned that $d_{ij}^{(t)}\le r$, it is uniformly distributed in $[-r,r]$ because Bayes' rule preserves the uniformity when conditioning on $|Y(t)|\le r$. Therefore, if the coordinate difference starts at $Y(0)=y_0$ and $Y(t)$ has probability density $p_{t,y_0}(Y(t) = y)$, we can write
\begin{equation}
\label{eq:integ}
R(s)=\frac{1}{2r}\int_{-r}^{r}\int_{-r}^r p_{s,y_0}(y)dydy_0.
\end{equation}
In Appendix~\ref{app:brownian_Rs_calculation}, we derive $R(s)\propto s^{-1/2}$ by Fourier expansion of $p_{s,y_0}(y)$. 

\subsubsection{L\'evy flights}
\label{sssec:levy}

In this section, we describe the calculation of $R(s)$ for dynamic RGGs with coordinates undergoing a L\'evy flight. A L\'evy flight is a stochastic process exhibiting heavy-tailed fluctuations in displacements on any given timescale, except in the special case corresponding to Brownian motion (see Sec.~\ref{sssec:brownian}). L\'evy flights have been used for describing, e.g., animal foraging patterns \cite{campeau2022evolutionary}, human mobility \cite{rhee2011levy}, and anomalous diffusion \cite{weeks1995observation}.

To describe L\'evy flights, we consider the Fourier transform $\mathcal{F}$ with positive sign convention, operating on functions $q(x)$, i.e., $\varphi(\omega)=[\mathcal{F}q](\omega)=\int e^{ix\omega}q(x)dx$; its inverse is $[\mathcal{F}^{-1}\varphi](x)=\frac{1}{2\pi}\int e^{-ix\omega}\varphi(\omega)d\omega=q(x)$. When $q(x)$ is a probability density, $\varphi(\omega)$ is called its {\it characteristic function}. For symmetric L\'evy flights $X(t)$ originating at $X(0)=x_0$ in continuous time $t\ge 0$, the probability density of $X(t)$, denoted $p_{t,x_0}(x)$, cannot be expressed in a simple closed form. However, its characteristic function $\varphi_{t,x_0}(\omega)$ is given by \cite{feller1991,ben2000diffusion,kyprianou2014fluctuations}
\begin{equation}
     \varphi_{t,x_0}(\omega)=e^{ix_0 \omega-t|\omega|^{\alpha}}.
\end{equation}
Therefore, 
\begin{equation}
p_{t,x_0}(x)=[\mathcal{F}^{-1}\varphi_{t,x_0}](x).
\end{equation}

The parameter $\alpha$ tunes tail behavior of the distribution of displacement $\delta=|x-x_0|$; it has density of the form $p(\delta)\propto \delta^{-(1+\alpha)}$ for $\alpha<2$, and hence diverging second moment, with heavier-tailed displacements at smaller $\alpha$. The quantity $t^{1/\alpha}$ is a {\it scale parameter} for $p_{x_0,t}(x)$, in that $X(t)/t^{1/\alpha}$ and $X(1)$ have the same distribution for all $t>0$.

\begin{figure}[h]
    \centering
\includegraphics[scale=0.45,trim = 620 430 0 150,clip]{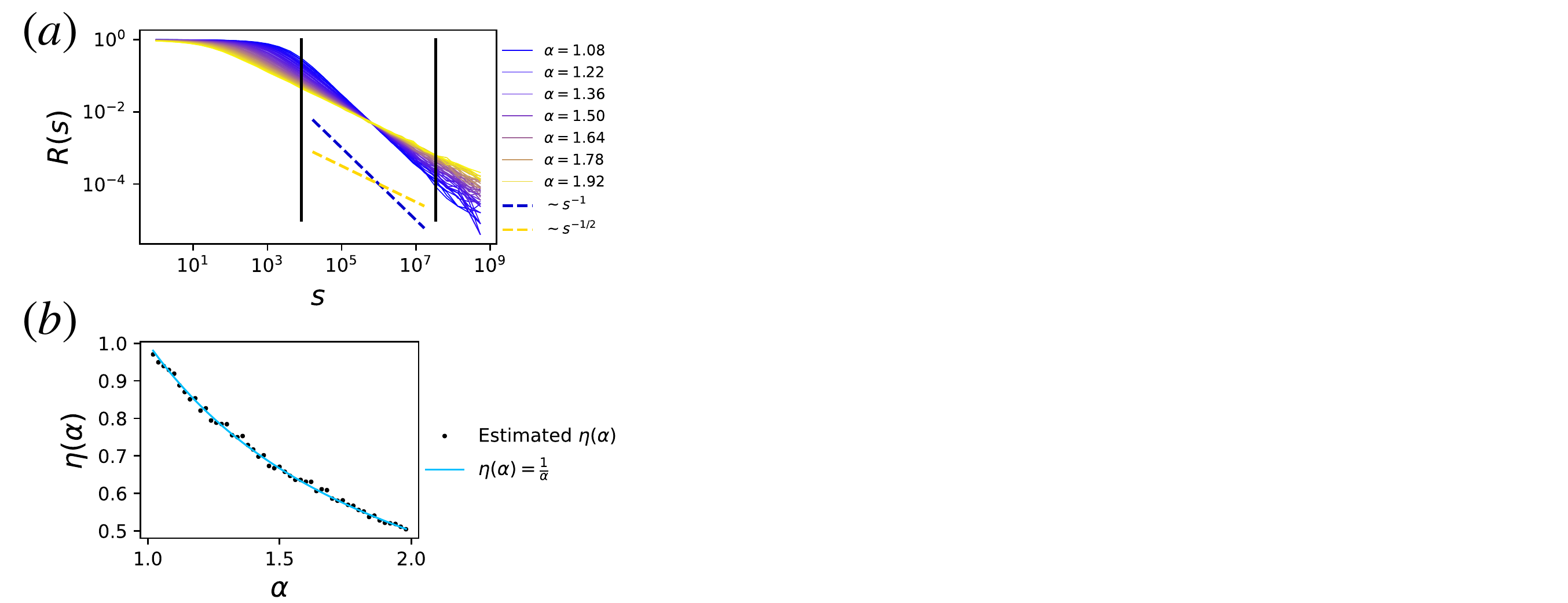}
    \caption{$R(s)$ for dynamic RGGs with L\'evy flights governing the coordinates of the nodes. (a) $R(s)$ obtained in simulations of L\'evy flights on the real line (corresponding to $n=\infty$). For node-pairs whose coordinates are initially uniform under the constraint that they are adjacent to each other, $R(s)$ is computed by sampling coordinates at a sequence of exponentially growing discrete times $t\in\{2^k\}_{k=0}^{29}$, mapped from continuous time via $\tilde{t}=(t-1)\tau$ with $\tau= 10^{-6}$. For visual reference, the dashed lines display $s^{-1/\alpha}$ at $\alpha=2$ and $\alpha=1$. For each $\alpha$ value, we measured the slope of the linear regression between $\ln R(s)$ and $\ln s$, denoted by $-\eta(\alpha)$, within the range of $s$ indicated by the two vertical black lines, with the lower one selected to avoid the plateau regions at small $s$ and the upper one selected to avoid large fluctuations at large $s$. (b) $\eta(\alpha)$ from the numerical data shown in (a). There is good agreement between the theoretical prediction, i.e., $\eta(\alpha)=1/\alpha$ (shown by the solid line), and the simulation results (shown by the dots).}
    \label{fig:fits_levy_corrected}
\end{figure}

By analyzing the asymptotics of Eq.~\eqref{eq:integ}, we show in Appendix~\ref{app:eta_Rs_alpha} that
\begin{equation}
\label{eq:Rs_levy_asymptotic}
R(s) \propto s^{-1/\alpha}
\end{equation}
for $\alpha\in(1,2]$. At $\alpha=2$, this agrees with the result for the Brownian motion (Sec.~\ref{sssec:brownian}). See Fig.~\ref{fig:fits_levy_corrected}a for visualization of $R(s)$ in simulations across $\alpha\in [1,2]$ and Fig.~\ref{fig:fits_levy_corrected}b for numerical validation of Eq.~\eqref{eq:Rs_levy_asymptotic}.

\subsubsection{Dynamic metapopulation proximity networks}
\label{sssec:dynamic_meta}

In this section, we describe the calculation of $R(s)$ for the dynamic metapopulation proximity network described in Sec~\ref{ssec:metapop}. In this case, $R(s)$ is the probability that two random walkers are on a common site at time $t+s$ conditioned on them having been on a common site at time $t$. In other words, we obtain
\begin{equation}
\label{eq:Rs_meta_expression}
    R(s)=\mathbb{P}\left(\left. X_i^{(t+s)}=X_j^{(t+s)}\right\vert X_i^{(t)}=X_j^{(t)}\right).
\end{equation}

\begin{figure}[h]
    \centering
    \includegraphics[scale=0.18,trim = 320 340 0 130,clip]{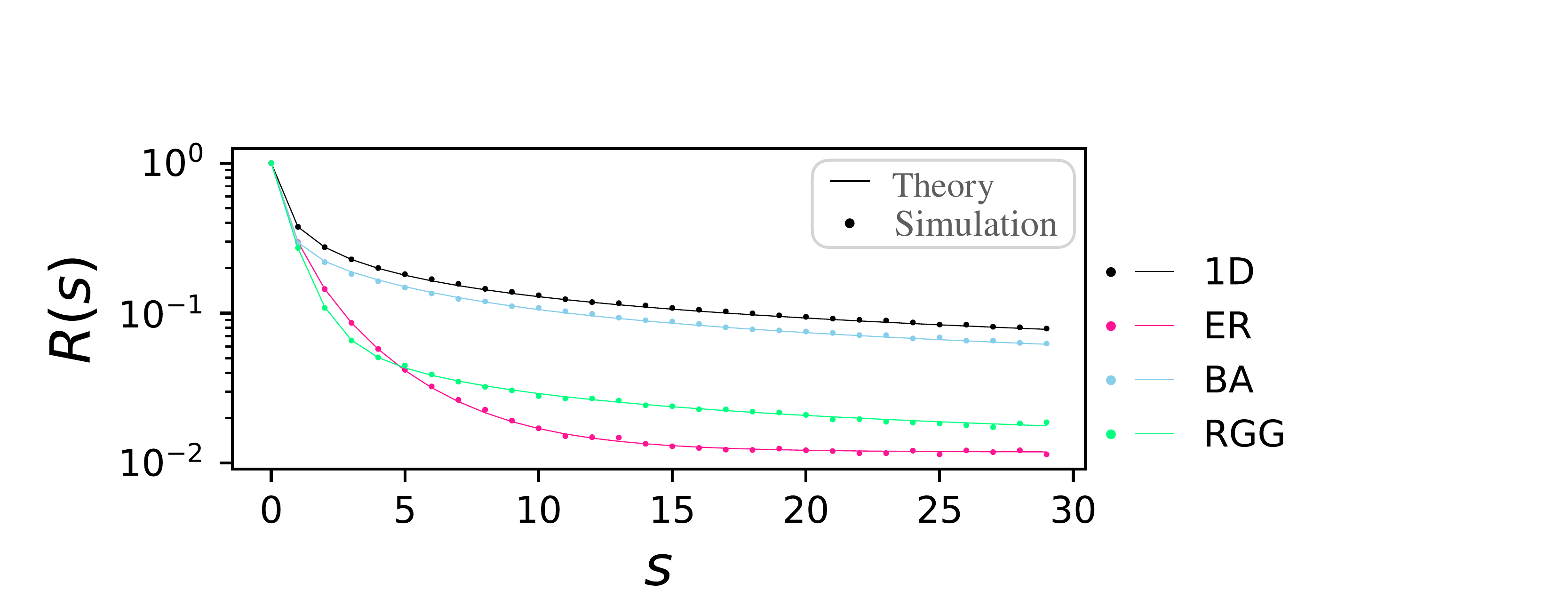}
    \caption{$R(s)$ for dynamic metapopulation proximity networks. The lines represent theoretical results. The circles represent numerical results.
 We computed the theoretical $R(s)$ using Eq.~\eqref{eq:Rs_metapopulation}. We compared the theoretical and numerical results for the following four
metapopulation graphs: (i) a one-dimensional lattice (i.e., chain graph) with $N=100$ nodes, (ii) an Erd\H{o}s-R\'enyi $G(N,p)$ graph with $N=100$ and $p=0.05$, (iii) a sample from the Barab\'asi-Albert model \cite{barabasi1999emergence} with $N=200$ nodes and $m=1$ edge carried by each newly arriving node (yielding a tree), initialized as a single node, and (iv) a random geometric graph formed by sprinkling $N=100$ sites into the two-dimensional unit square, connecting all pairs whose Euclidean distance is at most $0.15$, without periodic boundary conditions, and conditioned on connectedness of the entire network.
}
    \label{fig:Rs_metapop}
\end{figure}

As we show in Appendix~\ref{app:metapopulation_graph}, we obtain
\begin{equation}
\label{eq:Rs_metapopulation}
R(s)=\frac{\mathbf{1}^{\top} Q_s\mathbf{1}}{N\langle k^2\rangle},
\end{equation}
where $Q_s$ is an $N\times N$ matrix determined by the structure of the metapopulation graph and the transition probability of the random walk, $\langle k^2\rangle=\frac{1}{N}\sum_{I=1}^Nk_I^2$ is the second moment of the degree distribution of the metapopulation graph, $\mathbf{1}$ is the $N$-dimensional column vector with ones, and $^{\top}$ represents the transposition. The matrix $Q_s$ is defined elementwise by $(Q_s)_{IJ}=((W^s)_{IJ}k_J)^2$ for $I, J \in \{1, \ldots, N\}$, where $W^s$ is the $s$th power of the transition probability matrix of the random walk. See Fig.~\ref{fig:Rs_metapop} for comparison of $R(s)$ between theory and simulation for several metapopulation graphs; in all cases, the simulation and theoretical results match reasonably well. In Appendix~\ref{app:metapopulation_graph}, we re-express Eq.~\eqref{eq:Rs_metapopulation} as a sum of different exponentially decaying contributions as
\begin{equation}
R(s)=\frac{1}{N\langle k^2 \rangle} \sum_{i=1}^N k_i^2 \sum_{j=1}^N
\left[ \sum_{\ell=1}^N e^{-\psi_\ell s} \left( \boldsymbol{u}^{\text{R}}_{\ell} \right)_i \left( \boldsymbol{u}^{\text{L}}_{\ell} \right)_j \right]^2,
\end{equation}
where $\psi_\ell=-\log(1-q + q \lambda_{\ell})$; $\lambda_{\ell}$ is the $\ell$th eigenvalue of a symmetric matrix associated with the adjacency matrix of the metapopulation graph; $\boldsymbol{u}_\ell^R$ and $\boldsymbol{u}_\ell^L$ are the $\ell$th right and left eigenvectors of $Q_s$.

\subsubsection{Dynamic AND model}

\begin{figure}[h]
    \centering
    \includegraphics[scale=0.18,trim = 300 230 0 100, clip]{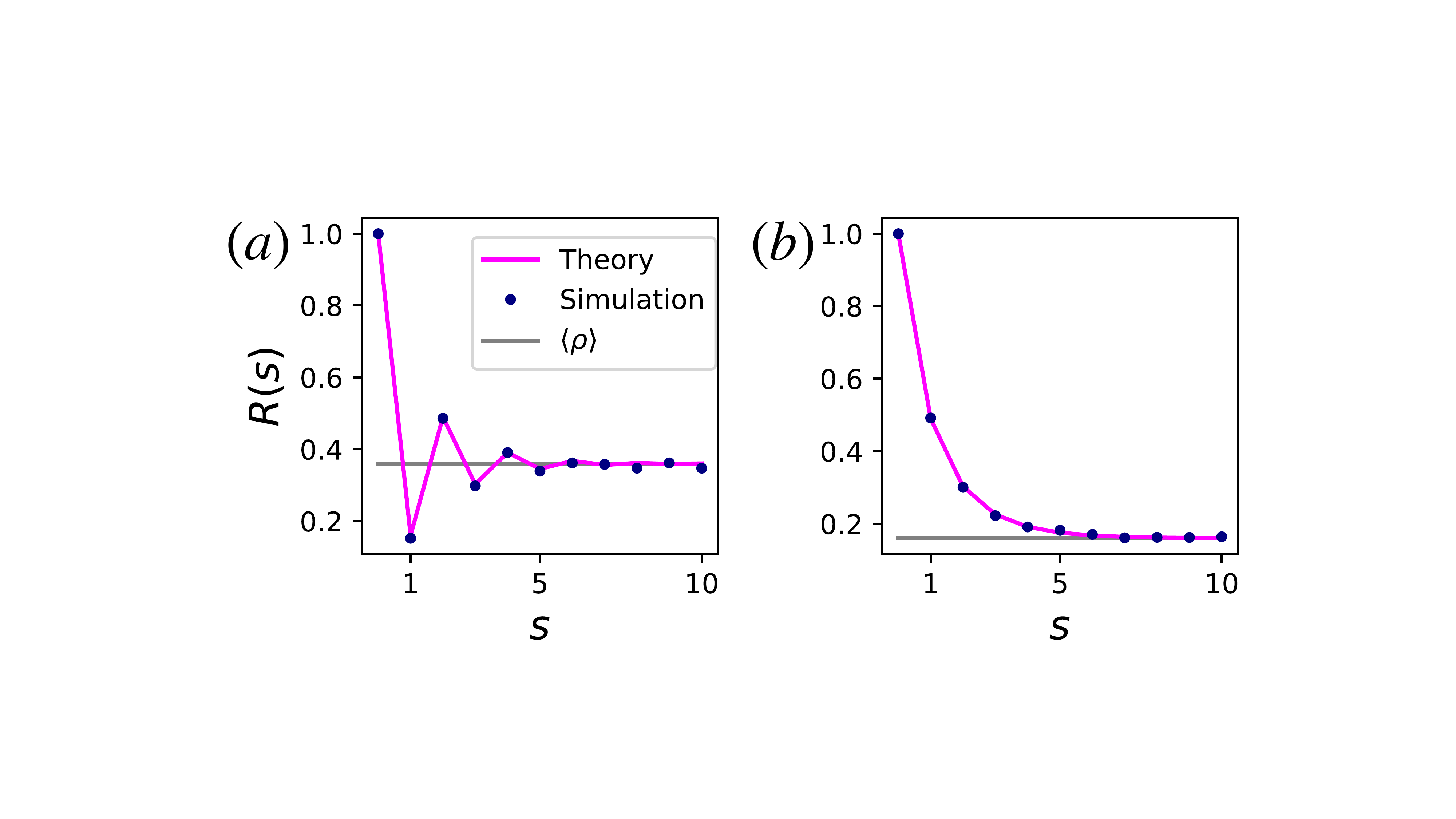}
    \caption{$R(s)$ in the discrete-time AND model. (a) $R(s)$ for the state-transition probabilities satisfying $P_{\ell h}+P_{h\ell}>1$ (specifically, $P_{\ell h} = 0.9$ and $P_{h \ell} = 0.6$), yielding oscillations. (b) $R(s)$ when $P_{\ell h}+P_{h\ell}<1$ (specifically, $P_{\ell h} = 0.2$ and $P_{h \ell} = 0.3$), yielding no oscillations. In both cases, the numerical results match reasonably well with the theoretical result given by Eq.~\eqref{eq:R(s)-final-AND-model}. The simulation results were from all pairs of $n=100$ nodes, simulated for $50$ time steps.}
    \label{fig:Rs_AND}
\end{figure}

In the discrete-time dynamic AND model (Sec.~\ref{ssec:dynamic_and}), the persistence $R(s)$ is the probability that both of the two nodes that are initially in the $h$ state are also in the $h$ state after $s$ time steps. Therefore, we obtain
\begin{equation}
\label{eq:Rs_AND_expression}
    R(s)=\mathbb{P}\left(\left. X_i^{(t+s)}=X_j^{(t+s)}=h\right\vert X_i^{(t)}=X_j^{(t)}=h\right).
\end{equation}
In Appendix~\ref{app:ANDmodel}, we analyze the two-state Markov chain that each node independently undergoes to obtain
\begin{equation}
    R(s)=\langle\rho\rangle+2\sqrt{\langle\rho\rangle}\left(1-\sqrt{\langle\rho\rangle}\right)\zeta^s+\left(1-\sqrt{\langle\rho\rangle}\right)^2\zeta^{2s},
\label{eq:R(s)-final-AND-model}
\end{equation}
where
\begin{equation}
\zeta=1-P_{\ell h}-P_{h \ell} \in (-1, 1).
\label{eq:def-zeta}
\end{equation}
Recall, $P_{\ell h}$ and $P_{h \ell}$ are the $\ell\rightarrow h$ transition probability and the $h\rightarrow \ell$ transition probability, respectively, in one step of the Markov chain. 
As expected, Eq.~\eqref{eq:R(s)-final-AND-model} yields $R(0)=1$ and $\lim_{s\to\infty}R(s)=\langle \rho\rangle$. 
The asymptotic behavior of $R(s)-\langle \rho\rangle$ is governed by the $\zeta^s$ term on the right-hand side of Eq.~\eqref{eq:R(s)-final-AND-model}.
See Fig.~\ref{fig:Rs_AND} for comparison between theory and simulation both for $\zeta<0$, yielding oscillatory relaxation,  and for $\zeta>0$, yielding relaxation without oscillation.

\section{Discussion}
\label{sec:discussion}

We have explored the long-timescale autocorrelation patterns of edges in temporal networks as quantified by the persistence $R(s)$, i.e., the probability of an edge existing after time $s$ given that the same edge currently exists, in several temporal network models. We have particularly elucidated the relationship between {\it dynamic} node variables $X^{(t)}$ and {\it dynamic} network structure $A^{(t)}$. We have considered the idealized deterministic dependency of the form $A^{(t)}=F(X^{(t)})$, e.g., establishing edges via pairwise proximity (Sec.~\ref{ssec:dynamic_RGGs}) or via co-location at a metapopulation site (Sec.~\ref{ssec:metapop}). As a result of the model specifications including the node-variable space $\mathcal{X}$, the mechanism of node-variable dynamics, and the mapping $F$, we find a wide variety of possible steady-state autocorrelation patterns as quantified by $R(s)$. Autocorrelation is one of several classical metrics in time series analysis and dynamical systems theory with application to temporal networks \cite{lacasa2022correlations,caligiuri2023lyapunov,papaefthymiou2024fundamental,jo2024temporal}.

This study is built on previous studies of network models with latent node variables, and in particular, dynamic latent-space models \cite{sewell2015latent,kim2018review} and temporal hidden-variable models \cite{hartle2021dynamic}. Many such network models have randomness at the network-structural level, and the question of how node-variable dynamics influences $R(s)$ is equally applicable in such stochastic settings. For instance, in the activity-driven model (ADM) of temporal networks \cite{perra2012activity}, random node variables (activities) influence connectivity in a stochastic manner. The original ADM and many variants of it have {\it static} random node activities, but several model variants involve {\it dynamic} activity values \cite{zino2018modeling,sheng2023constructing}; in such models, it would be interesting to examine how the chosen stochastic process governing node activities affects $R(s)$.

There are many further opportunities to explore $R(s)$ in related models. For instance, to yield patterns such as cutoffs of $R(s)$ at large $s$, power-law tails $\propto s^{-\eta}$ for $\eta\not\in[1/2,1]$ \cite{jo2024temporal}, and patterns of periodic oscillations across scales \cite{clauset2012persistence} are outstanding questions. A variety of autocorrelation patterns have been observed in real-world temporal network data \cite{lacasa2022correlations}. We considered simple node-variable dynamics and connection rules allowing analytical and numerical tractability, but many other options exist, e.g., those considered in communications systems engineering \cite{ye2010optimal,roy2011handbook} and human mobility science \cite{barbosa2018human,alessandretti2020scales,edsberg2022understanding}. Examples include the random waypoint model \cite{navidi2004stationary,bettstetter2004stochastic,hyytia2006spatial}, the gravity model of migration \cite{vanderkamp1977gravity,beine2016practitioners}, and models of correlated group mobility \cite{camp2002survey}. The question of how $R(s)$ behaves in dynamic RGGs with coordinates governed by such processes is generally open. We hope that this work inspires further modeling studies pertaining to dynamic networks influenced by dynamic node variables.

\section{Acknowledgements}

H.H. acknowledges support from the Santa Fe Institute. N.M. is supported in part by the National Science Foundation or NSF under grant DMS-2052720 and DMS-2204936, in part by JSPS KAKENHI under grants JP21H04595, 23H03414, 24K14840, and W24K030130, and in part by Japan Science and Technology Agency (JST) under grant JPMJMS2021.

\appendix

\section{Simulation methodology for Brownian motion and L\'evy flights}
\label{app:simulation}

To simulate dynamic RRGs with the node coordinates being governed by Brownian motion and L\'evy flights, we use the fact that these are {\it stable} processes \cite{feller1991vol2}, allowing us to efficiently sample their trajectories. Namely, rather than simulating a sequence of displacements occurring over time windows of fixed duration $\tau$, we construct trajectories sampled at exponentially increasing time intervals. This allows efficient computation of the resulting dynamical networks across many orders of magnitude of timescale.

We recall that we run these processes in continuous time $\tilde{t}$ and map $\tilde{t}\in\{0,\tau, \ldots,(T-1)\tau\}$ to discrete time indices $t\in\{1,2,...,T\}$. By using the stability property, we may jump ahead from any $t_j \in\{1, \ldots, T-1\}$ to any other larger $t_{j+1} \in\{2, \ldots, T\}$ by sampling a displacement corresponding to duration $(t_{j+1} - t_j)\tau$. In so doing, we may select a sequence of times $t_j$ at which we measure the node variables and edges. We use $t_j$ that grows exponentially as $t_j=\lfloor t_0 \beta^{j-1}\rfloor$, where $t_0>0$, $\beta>1$, $j\in\{1, \ldots, j_+\}$, $j_+$ is the number of times considered, and $\lfloor x \rfloor$ is the greatest integer that is less than or equal to $x$. We can then simulate a coordinate trajectory in $O(j_+)$ time complexity, as opposed to $O(T)$. This sampling procedure allows us to efficiently simulate long-timescale dynamics of Brownian walks and L\'{e}vy flights, and accordingly, to efficiently generate network trajectories.

\section{Calculation of $R(s)$ for Brownian motion of coordinates}
\label{app:brownian_Rs_calculation}

In this section, we derive $R(s)\propto s^{-1/2}$ for one-dimensional dynamic RGGs under Brownian motion. We denote by $p_{K}(x',t'|x,t)$ the normal density describing the position $x'$ of a Brownian walker after duration $t'-t$ ($>0$) starting from a point mass at $x$, i.e., $p_{K}(x',t|x,t)=\delta(x-x')$; it has mean $x$, variance $K(t'-t)$, and probability density function
\begin{equation}
p_{K}(x',t'|x,t)=\frac{1}{\sqrt{2\pi K(t'-t)}}\exp\left[ -\frac{(x'-x)^2}{K(t'-t)}\right].
\end{equation}

Then, the displacement between two walkers $Y(t)=X_i(t)-X_j(t)$ diffuses as $p_{2K}(y',t'|y,t)$. Conditioned that two nodes are initially adjacent to each other, $Y(0)$ is uniformly distributed on $[-r,r]$ as discussed in Sec.~\ref{sssec:brownian}. The persistence $R(s)$ is therefore
\begin{equation}
\begin{aligned}
R(s)&=\frac{1}{2r}\int_{-r}^{r}\int_{-r}^{r}p_{2K}(s,z'|0,z)dzdz'.\\
\end{aligned}
\label{eq:R(s)-Brownian-integral-appendix}
\end{equation}
The associated Gaussian density is
\begin{equation}
    p_{2K}(s,z'|0,z) = \frac{1}{2\sqrt{\pi Ks}}\exp\left[-\frac{(z'-z)^2}{2Ks}\right],
\label{eq:Gaussian}
\end{equation}    
By substituting Eq.~\eqref{eq:Gaussian} in Eq.~\eqref{eq:R(s)-Brownian-integral-appendix} and using $e^x=\sum_{a=0}^{\infty}x^a/a!$, we obtain
\begin{align}\label{eq:Rs_approx}
R(s)&= \frac{1}{4r\sqrt{\pi Ks}}\sum_{a=0}^\infty\left[-\frac{(z'-z)^2}{2Ks}\right]^a\frac{1}{a!} \notag\\
&=\frac{s^{-1/2}}{4r\sqrt{\pi K}}+O(s^{-3/2}).
\end{align}
Therefore, we have $R(s)\propto s^{-1/2}$ as the leading contribution for $s\gg 1$.

\section{Calculation of $R(s)$ for L\'evy flights}
\label{app:eta_Rs_alpha}

In this section, we compute $R(s)$ in dynamic RGGs when the node's coordinate undergoes a L\'evy flight. The derivation follows similar logic as in the case of Brownian motion (Appendix~\ref{app:brownian_Rs_calculation}). Recall that $\alpha$-stable distributions have a probability density function that is not expressible in closed form; rather, its characteristic function is. The characteristic function for the displacement $\varphi_{Y(t)}(\omega):=\langle e^{iY(t)\omega }\rangle$ is related to that for the individual positions $\varphi_{X_i(t)}(\omega):=\langle e^{iX_i(t)\omega}\rangle$ and $\varphi_{X_j(t)}(\omega):=\langle e^{iX_j(t)\omega }\rangle$ by the fact that the characteristic function of a weighted sum of random variables is a product of the characteristic functions with arguments shifted by the coefficients of the sum. Namely, since $Y(t)=X_i(t)+(-1)X_j(t)$, we have
\begin{equation}
\begin{aligned}
\varphi_{Y(t)}(\omega)&=\varphi_{X_i(t)}(z)\varphi_{X_j(t)}(-\omega)\\
&=e^{i\left[ X_i(0)-X_j(0) \right] \omega-2t|\omega|^{\alpha}},
\end{aligned}
\end{equation}

which is in the form of the characteristic function of a L\'evy flight with different parameters. Thus, the displacement $Y(t)$ also undergoes an $\alpha$-stable process, of mean $X_i(0)-X_j(0)=Y(0)$, and scale parameter $(2t)^{1/\alpha}$ rather than $t^{1/\alpha}$. Using the uniformity of the initial displacement density over $[-r,r]$ (see Appendix~\ref{app:brownian_Rs_calculation}) and $p(Y(s),s|Y(0),0)$, defined as the probability density of displacement $Y(s)$ at time $s$, given the initial displacement $Y(0)$ at time $0$, e obtain
\begin{align}
R(s)&=\mathbb{P}(Y(s)\in[-r,r]|Y(0)\in[-r,r]) \notag\\
&=\frac{1}{2r}\int_{-r}^r\int_{-r}^r p(Y(s),s|Y(0),0)dY(0)dY(s).
\label{eq:R(s)-Levy-stable-1}
\end{align}
We represent the transition probability density $p(Y(s),s | Y(0),0)$ in terms of the inverse Fourier transformation of its characteristic function, $\varphi_{Y(t)}$, as follows:
\begin{equation}
p(Y(s),s | Y(0),0)=\frac{1}{2\pi}\int_{-\infty}^{\infty} e^{-iz(Y(s)+Y(0))-2s|z|^{\alpha}}dz.
\label{eq:inverse-Fourier-Levy-stable}
\end{equation}
By substituting Eq.~\eqref{eq:inverse-Fourier-Levy-stable} in Eq.~\eqref{eq:R(s)-Levy-stable-1}, we obtain
\begin{equation}
\begin{aligned}
\label{eq:intermediate}
R(s)&=\frac{1}{4\pi r}\int_{-r}^r\int_{-r}^r\int_{-\infty}^{\infty} e^{-iz(y+y_0)-2s|z|^{\alpha}}dzdydy_0  \\
&=\frac{1}{4\pi r}\int_{-\infty}^{\infty}\left(\int_{-r}^re^{-izu}du\right)^2e^{-2s|z|^\alpha}dz  \\
&=\frac{1}{\pi}\int_{-\infty}^\infty\left[\frac{\sin(zr)}{zr}\right]^2e^{-2r^{-1}s|zr|^\alpha}d(zr)  \\
&=\frac{2}{\pi}\int_{0}^{\infty}\mathrm{sinc}(w)^2e^{-\upsilon w^\alpha}dw,
\end{aligned}
\end{equation}
where
\begin{equation}
\mathrm{sinc}(w) \equiv \frac{\sin(w)}{w}
\end{equation}
and
\begin{equation}
\upsilon =\frac{2 s}{r}.
\label{eq:def-upsilon}
\end{equation}
The function $\mathrm{sinc}(w)$ has a removable singularity at zero and has the following Taylor expansion:
\begin{align}
\mathrm{sinc}(w)&=\sum_{n=0}^\infty \frac{(-1)^n w^{2n}}{(2n+1)!} \notag\\
&= 1 - \frac{w^2}{3!} + \frac{w^4}{5!} +O(w^6).
\label{eq:sinc-Taylor}
\end{align}
Equation~\eqref{eq:sinc-Taylor} yields
\begin{align}
\mathrm{sinc}(w)^2&=\left[\sum_{n=0}^\infty \frac{(-1)^n w^{2n}}{(2n+1)!}\right]^2 \notag\\
&=\sum_{n=0}^\infty \sum_{q=0}^\infty \frac{(-1)^{n+q} w^{2(n+q)}}{(2n+1)!(2q+1)!} \notag\\
&=\left[ 1 - \frac{w^2}{3!} +\frac{w^4}{5!} +O(w^6)\right]^2 \notag\\
&=1-\frac{w^2}{3}+\frac{2w^4}{45}+O(w^6).
\end{align}
Therefore, we obtain
\begin{equation}
R(s) = \frac{2}{\pi}\sum_{n=0}^\infty \sum_{q=0}^\infty \frac{(-1)^{n+q}}{(2n+1)!(2q+1)!}\int_0^{\infty}w^{2(n+q)}e^{-\upsilon w^\alpha}dw.
\label{eq:R(s)-alpha-stable-2}
\end{equation}
We now use the formula
\begin{equation}
\int_0^\infty x^c e^{-ax^b} dx = \frac{1}{b}\ a^{-\frac{c+1}{b}}\Gamma\left(\frac{c+1}{b}\right),
\label{eq:gamma-formula}
\end{equation}
where $\Gamma(z)=\int_0^\infty t^{z-1}e^{-t}dt$ is the gamma function. By substituting $c=2(n+q)$, $a=\upsilon$, and $b=\alpha$ in Eq.~\eqref{eq:gamma-formula}, we obtain
\begin{equation}
\int_{0}^{\infty}w^{2(n+q)}e^{-\upsilon w^{\alpha}}dw=\frac{1}{\alpha} \upsilon^{-\frac{2(n+q)+1}{\alpha}}\Gamma\left(\frac{2(n+q)+1}{\alpha}\right).
\label{eq:int-with-gamma-1}
\end{equation}
By substituting Eq.~\eqref{eq:int-with-gamma-1} in Eq.~\eqref{eq:R(s)-alpha-stable-2}, we obtain
\begin{align}
R(s)&=\frac{2\upsilon^{-\frac{1}{\alpha}}}{\alpha \pi}\sum_{n=0}^\infty \sum_{q=0}^\infty \frac{(-1)^{n+q}\Gamma\left(\frac{2(n+q)+1}{\alpha}\right)}{(2n+1)!(2q+1)!} \upsilon^{-\frac{2(n+q)}{\alpha}} \notag\\
&=\frac{2\upsilon^{-\frac{1}{\alpha}}}{\alpha \pi}\left[\Gamma\left(\frac{1}{\alpha}\right)-\frac{\Gamma\left(\frac{3}{\alpha}\right)}{18}\upsilon^{-\frac{2}{\alpha}}+O\left(\upsilon^{-\frac{4}{\alpha}}\right)\right].
\label{eq:R(s)-alpha-stable-3}
\end{align}
By substituting Eq.~\eqref{eq:def-upsilon} in Eq.~\eqref{eq:R(s)-alpha-stable-3}, we obtain
\begin{widetext}
\begin{align}
\label{eq:Rs_leading_scaling}
R(s)&=\frac{2}{\alpha\pi}\left(\frac{r}{2}\right)^{\frac{1}{\alpha}}s^{-\frac{1}{\alpha}}\left[\Gamma\left(\frac{1}{\alpha}\right)-\frac{\Gamma\left(\frac{3}{\alpha}\right)}{18}\left(\frac{r}{2}\right)^{\frac{2}{\alpha}}s^{-\frac{2}{\alpha}}+O\left(s^{-\frac{4}{\alpha}}\right)\right] \notag\\
&=C_1(\alpha,r)s^{-\frac{1}{\alpha}}-C_2(\alpha,r)s^{-\frac{3}{\alpha}}+O\left(s^{-\frac{5}{\alpha}}\right),
\end{align}
\end{widetext}
where $C_1(\alpha,r)$ and $C_2(\alpha,r)$ are quantities independent of $s$. Therefore, we obtain $R(s)\propto s^{-\eta(\alpha)}$ with $\eta(\alpha) = \alpha^{-1}$ for $s\gg 1$.

\section{Probability distribution for the location of a pair of walkers in metapopulation graphs}
\label{app:P_I_given_x_eq_x}

Let $(x,x')$ represent locations of two independent random walkers on the metapopulation graph. In this section, we evaluate $\mathbb{P}(x'=I|x=x')$, where $I\in \{1, \ldots, N \}$ is a site of the metapopulation graph. Using Bayes' theorem, we obtain
\begin{equation}
\mathbb{P}(x'=I|x'=x)=\frac{\mathbb{P}(x'=x|x'=I)\mathbb{P}(x'=I)}{\mathbb{P}(x'=x)}.
\end{equation}
Because of the independence of the two random walkers, assuming stationarity, we obtain
\begin{align}
\mathbb{P}(x'=x|x'=I)&=\mathbb{P}(x=I|x'=I) \notag\\
&=\mathbb{P}(x=I) \notag\\
&=\frac{k_I}{\sum_{J=1}^Nk_J},
\label{eq:metapop-a}
\end{align}
where we remind that $k_I$ is the degree of the $I$th node in the metapopulation graph. Note that $\sum_{J=1}^Nk_J=2M$, with $M$ being the number of edges in the metapopulation graph. Using Eq.~\eqref{eq:metapop-a} and
\begin{equation}
\mathbb{P}(x=I)=\mathbb{P}(x'=I)= \frac{k_I}{N\langle k\rangle},
\end{equation}
where $\langle k\rangle=2M/N$ is the average degree of the metapopulation graph, we obtain
\begin{align}
\label{eq:C3}
\mathbb{P}(x'=I|x'=x)&=\frac{\mathbb{P}(x'=x|x'=I)\mathbb{P}(x'=I)}{\mathbb{P}(x'=x)} \notag\\
&=\frac{\mathbb{P}(x=I)^2}{\mathbb{P}(x'=x)} \notag\\
&=\frac{k_I^2}{(N\langle k\rangle)^2\mathbb{P}(x'=x)},
\end{align}
Using the normalization condition, $\sum_{I=1}^N \mathbb{P}(x'=I|x'=x) = 1$, we obtain
\begin{align}
\label{eq:C4}
\mathbb{P}(x'=x)&=\sum_{I=1}^N\frac{k_I^2}{(N\langle k \rangle)^2} \notag\\
&=\frac{\langle k^2\rangle}{N\langle k\rangle^2},
\end{align}
where $\langle k^2\rangle=N^{-1}\sum_{I=1}^Nk_I^2$ is the second moment of the degree distribution of the metapopulation graph.

By combining Eqs.~\eqref{eq:C3} and \eqref{eq:C4}, we obtain
\begin{align}
\mathbb{P}(x'=I|x=x')&=\frac{k_I^2}{(N\langle k\rangle)^2}\left/\frac{N\langle k\rangle^2}{\langle k^2\rangle}\right. \notag\\
&=\frac{k_I^2}{N\langle k^2\rangle}.
\label{eq:joint-visit-meta-final}
\end{align}
Hence conditioning on {\it both} nodes being present at node $I$ results in a quadratic rather than linear dependence on $k_I$.

\section{Calculation of $R(s)$ for dynamic metapopulation proximity networks}
\label{app:metapopulation_graph}

In this section, we analyze $R(s)$ in the dynamic metapopulation proximity network. We recall that two random walkers $i$ and $j$ are adjacent at any given discrete time $t$ if and only if they visit the same site.

Let $p(J,t';I,t)$ denote the probability that a random walker resides on site $J$ at time $t'$ given that it resided on site $I$ at time $t<t'$. The probability that a pair of walkers, $i$ and $j$, reside on a common site after $s$ time steps given that they both are initialized on site $I$ is given by
\begin{widetext}
\begin{equation}
\label{eq:prob_given_I_meta}
\mathbb{P}\left(X_i^{(t+s)}=X_j^{(t+s)}\left\vert X_i^{(t)}=X_j^{(t)}=I\right.\right)=\sum_{J=1}^Np(J,t+s;I,t)^2.
\end{equation}
\end{widetext}
Note that $p(J,t+s;I,t)=p(J,s;t,0)$. The conditional probability given by Eq.~\eqref{eq:prob_given_I_meta} is related to but different from
\begin{align}
R(s)&=\mathbb{P}\left(A_{ij}^{(t+s)}=1\left\vert A_{ij}^{(t)}=1\right.\right) \notag\\
&=\mathbb{P}\left(X_i^{(t+s)}=X_j^{(t+s)}\left\vert X_i^{(t)}=X_j^{(t)}\right.\right).
\label{eq:R(s)-meta-1}
\end{align}
To connect Eqs.~\eqref{eq:prob_given_I_meta} and \eqref{eq:R(s)-meta-1}, we
first note that, for an arbitrary event $\mathcal{A}$, we obtain
\begin{align}
\label{eq:P_A_given_same_and_I}
& \mathbb{P}(\mathcal{A}|X_i^{(t)}=X_j^{(t)})\notag\\
=& \sum_{I=1}^N\mathbb{P}(\mathcal{A}|X_i^{(t)}=X_j^{(t)}, X_j^{(t)} = I)\mathbb{P}(X_j^{(t)} =I | X_i^{(t)}=X_j^{(t)})\notag\\
=& \frac{1}{N\langle k^2\rangle}\sum_{I=1}^N\mathbb{P}(\mathcal{A}|X_i^{(t)}=X_j^{(t)}, X_j^{(t)} = I)k_I^2,
\end{align}
where we used Eq.~\eqref{eq:joint-visit-meta-final} to derive the second equality.

By substituting the event $X_i^{(t+s)}=X_j^{(t+s)}$ in $\mathcal{A}$ in Eq.~\eqref{eq:P_A_given_same_and_I} and using
Eqs.~\eqref{eq:prob_given_I_meta} and \eqref{eq:R(s)-meta-1}, we obtain
\begin{align}
\label{eq:Rs_intermed_metapop}
R(s) &=\frac{1}{N\langle k^2\rangle}\sum_{I=1}^N
\mathbb{P}\left(X_i^{(t+s)}=X_j^{(t+s)}\left\vert X_i^{(t)}=X_j^{(t)}=I\right.\right) k_I^2 \notag\\
&=\frac{1}{N\langle k^2\rangle}\sum_{I=1}^N k_I^2\sum_{J=1}^Np(J,s;I,0)^2.
\end{align}

Let $\overline{A}^{\text{meta}} =(\overline{A}^{\text{meta}}_{IJ})_{1\le I<J\le N}$ with $\overline{A}^{\text{meta}}_{IJ}=\overline{A}^{\text{meta}}_{JI}$ be the adjacency matrix of an undirected metapopulation graph; $k_I=\sum_{J=1}^N \overline{A}^{\text{meta}}_{IJ}$. The transition probability matrix of the random walk
$W = (W_{IJ})$, where $W_{IJ}$ is the probability of moving from the $I$th to the $J$th site in a single time step, is given by
\begin{equation}
W = (1-q) I +q D^{-1} \overline{A}^{\text{meta}},
\label{eq:W}
\end{equation}
where $I$ is the $N\times N$ identity matrix, $D = \mathrm{diag}(k_1, \ldots, k_N)$, and $\mathrm{diag}$ denotes the diagonal matrix whose diagonal entries are given by its arguments, and $q$ is the probability that the agent moves to a neighboring site in one time step. Note that
\begin{equation}
p(J,s;I,0) = (W^s)_{IJ}.
\label{eq:connect-p-and-W-meta}
\end{equation}.

Now we write Eq.~\eqref{eq:W} as
\begin{equation}
W = D^{-1/2} \left[ (1-q) I + q \tilde{A}^{\text{meta}} \right] D^{1/2},
\label{eq:W-rewrite-sim}
\end{equation}
where $\tilde{A}^{\text{meta}} \equiv D^{-1/2} \overline{A}^{\text{meta}} D^{-1/2}$ is a symmetric matrix. Equation~\eqref{eq:W-rewrite-sim} implies that
$W$ and $(1-q) I + q\tilde{A}^{\text{meta}}$ are similar to each other (in the linear algebra sense). It is also the case that $(1-q) I + q \tilde{A}^{\text{meta}}$ and $\tilde{A}^{\text{meta}}$ share each eigenvector. Therefore, $\boldsymbol{u}_{\ell}^{\text{L}}$ and $\boldsymbol{u}_{\ell}^{\text{R}}$ given by Eqs.~(3.37) and (3.38) in \cite{masuda2020guide} are the left and right eigenvector, respectively, of $W$. The eigenvalues of $W$ are $1-q + q \lambda_{\ell}$, $\ell \in \{ 1, \ldots, N \}$, where $\lambda_{\ell}$ is the $\ell$th eigenvalue of $\tilde{A}^{\text{meta}}$.

So, similar to Eq. (3.41) in \cite{masuda2020guide}, we obtain
\begin{align}
W^s = \sum_{\ell=1}^N (1-q + q \lambda_{\ell})^s \boldsymbol{u}_{\ell}^{\text{R}} \boldsymbol{u}_{\ell}^{\text{L}}.
\end{align}
Therefore, the persistence is given by
\begin{equation}
R(s) = \frac{1}{N\langle k^2 \rangle} \sum_{i=1}^N k_i^2 \sum_{j=1}^N
\left[ \sum_{\ell=1}^N (1-q + q \lambda_{\ell})^s \left( \boldsymbol{u}^{\text{R}}_{\ell} \right)_i \left( \boldsymbol{u}^{\text{L}}_{\ell} \right)_j \right]^2.
\label{eq:R_s}
\end{equation}

Let us first look at the limit $s\to\infty$. We obtain $\lambda_1 = 1$, which holds true because 
$(\sqrt{k_1}, \ldots, \sqrt{k_N})/\sqrt{N \langle k\rangle}$ is the normalized left eigenvector of $\tilde{A}^{\mathrm{meta}}$ associated with $\lambda_1 = 1$; it is straightforward to verify this.
Using Eqs.~(3.37) and (3.38) in \cite{masuda2020guide}, we obtain 
\begin{equation}
\boldsymbol{u}_1^{\text{L}} = \frac{1}{\sqrt{N \langle k\rangle}} (k_1, \ldots, k_N)
\label{eq:u_0^L}
\end{equation}
and 
\begin{equation}
\boldsymbol{u}_1^{\text{R}} =  \frac{1}{\sqrt{N \langle k\rangle}} \begin{pmatrix}1 \\ \vdots \\ 1 \end{pmatrix}.
\label{eq:u_0^R}
\end{equation}
By combining Eqs.~\eqref{eq:R_s}, \eqref{eq:u_0^L}, and \eqref{eq:u_0^R}, and using
$\lim_{s\to\infty} (1-q + q \lambda_{\ell})^s = 0$ for $\ell \in \{2, \ldots, N \}$, we obtain
\begin{equation}
\lim_{s \to \infty}R(s) = \frac{1}{N\langle k^2 \rangle} \sum_{I=1}^N k_I^2 \sum_{J=1}^N
1^2 \cdot k_J^2 = \frac{\langle k^2 \rangle}{N \langle k \rangle^2}.
\end{equation}
This agrees with the prediction $\lim_{s\rightarrow\infty}R(s) = \langle \rho\rangle$, where, for the metapopulation graph, 
\begin{align}
\langle\rho\rangle =& \mathbb{P}(X_i^{\ast}=X_j^{\ast}) \notag\\
=& \sum_{I=1}^N(p_I^{\ast})^2 \notag\\
=& \frac{\langle k^2\rangle}{N\langle k\rangle}.
\label{eq:<rho>-metapop-final}
\end{align}
In Eq.~\eqref{eq:<rho>-metapop-final}, the quantities with ${}^\ast$ represent those in stationarity, and we have used $p_I^\ast = k_I/N\langle k\rangle$.

By computing Eq.~\eqref{eq:R_s}, we obtain
\begin{widetext}
\begin{align}
R(s)=& \frac{\langle k^2 \rangle}{N \langle k \rangle^2} + \frac{2}{N^2 \langle k^2 \rangle \langle k \rangle}
\sum_{\ell=2}^N (1-q + q \lambda_{\ell})^s \cdot \left[ \sum_{i=1}^N k_i (\boldsymbol{u}_{\ell}^{\text{L}})_i \right]^2 \notag\\
& + \frac{\sum_{i=1}^N k_i^2 \sum_{j=1}^N \left[ \sum_{\ell=2}^N  
(1-q + q \lambda_{\ell})^s \left( \boldsymbol{u}_{\ell}^{\text{R}} \right)_i \left( \boldsymbol{u}_{\ell}^{\text{L}} \right)_j
\right]^2} {N \langle k^2 \rangle}.
\label{eq:R_s-2}
\end{align}
\end{widetext}
Under the assumption of widely spaced eigenvalues, $R(s)$ is governed by $(1-q + q \lambda_2)^s$. However, if $\lambda_2$, $\lambda_3$, $\ldots$, are not far from each other, Eq.~\eqref{eq:R_s-2} is sum of exponentials of different rates, and hence may visually resemble a subexponential decay over some scale of $s$ \cite{okada2020long,feldmann1998fitting,jiang2016two}.

One can obtain another expression of $R(s)$ by substituting Eq.~\eqref{eq:connect-p-and-W-meta} in Eq.~\eqref{eq:Rs_intermed_metapop} as follows:
\begin{align}
R(s) =& \frac{1}{N\langle k^2\rangle}\sum_{I=1}^N k_I^2\sum_{J=1}^N(W^s)_{JI}^2 \notag\\
=& \frac{1}{N\langle k^2\rangle}\sum_{I=1}^N \sum_{J=1}^N (k_I(W^s)_{JI})^2 \notag\\
=& \frac{1}{N\langle k^2\rangle}\sum_{I=1}^N \sum_{J=1}^N (Q_s)_{JI} \notag\\
=& \frac{\mathbf{1}^{\top} Q_s\mathbf{1}}{N\langle K^2\rangle},
\end{align}
where matrix $Q_s$ is defined by $(Q_s)_{IJ}=(k_J(W^s)_{IJ})^2$.

\section{Calculation of $R(s)$ for the AND model}
\label{app:ANDmodel}

In the AND model, the stationary probability that an arbitrary node is in state $h$ is equal to $P_{\ell h}/(P_{\ell h}+P_{h\ell})$. Because different nodes are assumed to behave independently and two nodes are adjacent if and only if both nodes are in  state $h$, the mean edge density is given by
\begin{equation}
\langle \rho\rangle=\left(\frac{P_{\ell h}}{P_{\ell h}+P_{h \ell}}\right)^2.
\end{equation}

For an arbitrary $i$th node, we define
\begin{equation}
y_s=\mathbb{P}(X_i^{(t+s)}=h|X_i^{(t)}=h).
\end{equation}
Probability $y_s$ obeys the following recursion:
\begin{align}
y_{s} &= y_{s-1}(1-P_{h\ell})+(1-y_{s-1})P_{\ell h} \notag\\
&=P_{\ell h}+(1-P_{h\ell}-P_{\ell h})y_{s-1}.
\label{eq:y_s-recursion-AND-model}
\end{align}
with the initial condition $y_0=1$. The solution of Eq.~\eqref{eq:y_s-recursion-AND-model} (see Appendix~\ref{app:recursion_solution} for the derivation) is
\begin{align}
\label{eq:ys}
y_s&=\frac{P_{\ell h}}{P_{h\ell}+P_{\ell h}}+\left(1-P_{h\ell}-P_{\ell h}\right)^s\frac{P_{h\ell}}{P_{h\ell}+P_{\ell h}} \notag\\
&=\sqrt{\langle \rho\rangle}+\left(1-\sqrt{\langle\rho\rangle}\right)\zeta^s,
\end{align}
where $\zeta$ is defined by Eq.~\eqref{eq:def-zeta}.
By exploiting the independence between different nodes, we obtain
\begin{align}
\label{eq:Rs_AND}
R(s) &=\mathbb{P}\left(X_i^{(t+s)}=X_j^{(t+s)}=h\left\vert X_i^{(t)}=X_j^{(t)}=h\right.\right) \notag\\
&=y_s^2 \notag\\
&=\left[\sqrt{\langle \rho\rangle}+\left(1-\sqrt{\langle\rho\rangle}\right)\zeta^s\right]^2 \notag\\
&=\langle\rho\rangle+2\sqrt{\langle\rho\rangle}\left(1-\sqrt{\langle\rho\rangle}\right)\zeta^s+\left(1-\sqrt{\langle\rho\rangle}\right)^2\zeta^{2s},
\end{align}
which is Eq.~\eqref{eq:R(s)-final-AND-model}.

\section{Solution to linear recursion}
\label{app:recursion_solution}

Consider the recursion given by
\begin{equation}
y_s=a+by_{s-1}.
\label{eq:y_s-recursion-general}
\end{equation}
Equation~\eqref{eq:y_s-recursion-general} yields
\begin{align}
y_s&=a+b(a+by_{s-2}) \notag\\
&=a+ab+b^2y_{s-2} \notag\\
&= \cdots \notag\\
&=b^sy_{0}+a\sum_{w=0}^{s-1}b^w \notag\\
&=b^sy_0+a\frac{1-b^s}{1-b}.
\label{eq:y_s-recursion-2}
\end{align}
In the case of the AND model, we substitute $a=P_{\ell h}$, $b=1-P_{h\ell}-P_{\ell h}$, and $y_0=1$ in Eq.~\eqref{eq:y_s-recursion-2} to obtain the first line of Eq.~\eqref{eq:ys}.


\begin{thebibliography}{10}


\bibitem{bianconi2001bose}
G.~Bianconi and A.-L. Barab{\'a}si.
\newblock Bose-Einstein condensation in complex networks.
\newblock {\em Physical Review Letters}, 86(24):5632, 2001.
\newblock \href {https://doi.org/10.1103/PhysRevLett.86.5632}
  {\path{doi:10.1103/PhysRevLett.86.5632}}.

\bibitem{caldarelli2002scale}
G.~Caldarelli, A.~Capocci, P.~De~Los~Rios, and M.~A. Mu\~{n}oz.
\newblock Scale-free networks from varying vertex intrinsic fitness.
\newblock {\em Physical Review Letters}, 89(25):258702, 2002.
\newblock \href {https://doi.org/10.1103/PhysRevLett.89.258702}
  {\path{doi:10.1103/PhysRevLett.89.258702}}.

\bibitem{boguna2003class}
M.~Bogu\~{n}{\'a} and R.~Pastor-Satorras.
\newblock Class of correlated random networks with hidden variables.
\newblock {\em Physical Review E}, 68(3):036112, 2003.
\newblock \href {https://doi.org/10.1103/PhysRevE.68.036112}
  {\path{doi:10.1103/PhysRevE.68.036112}}.

\bibitem{chung2003spectra}
F.~Chung, L.~Lu, and V.~Vu.
\newblock Spectra of random graphs with given expected degrees.
\newblock {\em Proceedings of the National Academy of Sciences of the United States of America},
  100(11):6313, 2003.
\newblock \href {https://doi.org/10.1080/15427951.2004.10129089}
  {\path{doi:10.1080/15427951.2004.10129089}}.

\bibitem{masuda2004analysis}
N.~Masuda, H.~Miwa, and N.~Konno.
\newblock {Analysis of scale-free networks based on a threshold graph with
  intrinsic vertex weights}.
\newblock {\em Physical Review E}, 70(3):036124, 2004.
\newblock \href {https://doi.org/10.1103/physreve.70.036124}
  {\path{doi:10.1103/physreve.70.036124}}.

\bibitem{penrose2003random}
M.~Penrose.
\newblock {\em Random Geometric Graphs}.
\newblock Oxford University Press, Oxford, 2003.
\newblock \href {https://doi.org/10.1093/acprof:oso/9780198506263.001.0001}
  {\path{doi:10.1093/acprof:oso/9780198506263.001.0001}}.

\bibitem{van2018sparse}
P.~van~der Hoorn, G.~Lippner, and D.~Krioukov.
\newblock Sparse maximum-entropy random graphs with a given power-law degree
  distribution.
\newblock {\em Journal of Statistical Physics}, 173:806, 2018.
\newblock \href {https://doi.org/10.1007/s10955-017-1887-7}
  {\path{doi:10.1007/s10955-017-1887-7}}.

\bibitem{serrano2008self}
M.~{\'A}. Serrano, D.~Krioukov, and M.~Bogu\~{n}{\'a}.
\newblock Self-similarity of complex networks and hidden metric spaces.
\newblock {\em Physical Review Letters}, 100(7):078701, 2008.
\newblock \href {https://doi.org/10.1103/physrevlett.100.078701}
  {\path{doi:10.1103/physrevlett.100.078701}}.

\bibitem{krioukov2010hyperbolic}
D.~Krioukov, F.~Papadopoulos, M.~Kitsak, A.~Vahdat, and M.~Bogu\~{n}{\'a}.
\newblock Hyperbolic geometry of complex networks.
\newblock {\em Physical Review E}, 82(3):036106, 2010.
\newblock \href {https://doi.org/10.1103/PhysRevE.82.036106}
  {\path{doi:10.1103/PhysRevE.82.036106}}.

\bibitem{karrer2011stochastic}
B.~Karrer and M.~E.~J. Newman.
\newblock Stochastic blockmodels and community structure in networks.
\newblock {\em Physical Review E}, 83(1):016107, 2011.
\newblock \href {https://doi.org/10.1103/PhysRevE.83.016107}
  {\path{doi:10.1103/PhysRevE.83.016107}}.

\bibitem{barucca2018disentangling}
P.~Barucca, F.~Lillo, P.~Mazzarisi, and D.~Tantari.
\newblock Disentangling group and link persistence in dynamic stochastic block
  models.
\newblock {\em Journal of Statistical Mechanics},
  2018(12):123407, 2018.
\newblock \href {https://doi.org/10.1088/1742-5468/aaeb44}
  {\path{doi:10.1088/1742-5468/aaeb44}}.

\bibitem{diaz2007dynamic}
J.~D\'iaz, D.~Mitsche, and X.~P\'erez.
\newblock Dynamic random geometric graphs.
\newblock {\em Preprint cs/0702074}, 2007.
\newblock URL: \url{https://arxiv.org/abs/cs/0702074}.

\bibitem{papaefthymiou2024fundamental}
E.~S. Papaefthymiou, C.~Iordanou, and F.~Papadopoulos.
\newblock Fundamental dynamics of popularity-similarity trajectories in real
  networks.
\newblock {\em Physical Review Letters}, 132(25):257401, 2024.
\newblock \href {https://doi.org/10.1103/PhysRevLett.132.257401}
  {\path{doi:10.1103/PhysRevLett.132.257401}}.

\bibitem{friel2016interlocking}
N.~Friel, R.~Rastelli, J.~Wyse, and A.~E. Raftery.
\newblock Interlocking directorates in irish companies using a latent space
  model for bipartite networks.
\newblock {\em Proceedings of the National Academy of Sciences of the United States of America},
  113(24):6629, 2016.
\newblock \href {https://doi.org/10.1073/pnas.1606295113}
  {\path{doi:10.1073/pnas.1606295113}}.

\bibitem{mazzarisi2020dynamic}
P.~Mazzarisi, P.~Barucca, F.~Lillo, and D.~Tantari.
\newblock A dynamic network model with persistent links and node-specific
  latent variables, with an application to the interbank market.
\newblock {\em European Journal of Operational Research}, 281(1):50, 2020.
\newblock \href {https://doi.org/10.1016/j.ejor.2019.07.024}
  {\path{doi:10.1016/j.ejor.2019.07.024}}.

\bibitem{ward2013gravity}
M.~D. Ward, J.~S. Ahlquist, and A.~Rozenas.
\newblock Gravity's rainbow: A dynamic latent space model for the world trade
  network.
\newblock {\em Network Science}, 1(1):95, 2013.
\newblock \href {https://doi.org/10.1017/nws.2013.1}
  {\path{doi:10.1017/nws.2013.1}}.

\bibitem{sewell2015latent}
D.~K. Sewell and Y.~Chen.
\newblock Latent space models for dynamic networks.
\newblock {\em Journal of the American Statistical Association},
  110(512):1646, 2015.
\newblock \href {https://doi.org/10.1080/01621459.2014.988214}
  {\path{doi:10.1080/01621459.2014.988214}}.

\bibitem{peres2013mobile}
Y. Peres, A. Sinclair, P. Sousi, and A. Stauffer,
\newblock Mobile geometric graphs: Detection, coverage and percolation,
\newblock {\em Probability Theory and Related Fields}, 156(1):273, 2013.
\newblock \href {https://doi.org/10.1007/s00440-012-0428-1}
  {\path{doi:10.1007/s00440-012-0428-1}}.

\bibitem{peixoto2023modelling}
T. P. Peixoto and M. Rosvall, 
\newblock Modelling temporal networks with Markov chains, community structures and change points, 
in \textit{Temporal Network Theory}, Springer, 2023, p. 65.
\newblock \href {https://doi.org/10.1007/978-3-030-23495-9_4}
  {\path{doi:10.1007/978-3-030-23495-9_4}}.
  

\bibitem{hartle2021dynamic}
H.~Hartle, F.~Papadopoulos, and D.~Krioukov.
\newblock Dynamic hidden-variable network models.
\newblock {\em Physical Review E}, 103(5):052307, 2021.
\newblock \href {https://doi.org/10.1103/PhysRevE.103.052307}
  {\path{doi:10.1103/PhysRevE.103.052307}}.

\bibitem{ghasemian2016detectability}
A.~Ghasemian, P.~Zhang, A~Clauset, C.~Moore, and L.~Peel.
\newblock Detectability thresholds and optimal algorithms for community
  structure in dynamic networks.
\newblock {\em Physical Review X}, 6(3):031005, 2016.
\newblock \href {https://doi.org/10.1103/physrevx.6.031005}
  {\path{doi:10.1103/physrevx.6.031005}}.

\bibitem{panisson2013fingerprinting}
A.~Panisson, L.~Gauvin, A.~Barrat, and C.~Cattuto.
\newblock Fingerprinting temporal networks of close-range human proximity.
\newblock In {\em 2013 IEEE International Conference on Pervasive Computing and
  Communications Workshops (PERCOM Workshops)}, p.~261. IEEE, 2013.
\newblock \href {https://doi.org/10.1109/percomw.2013.6529492}
  {\path{doi:10.1109/percomw.2013.6529492}}.

\bibitem{scholz2014predictability}
C.~Scholz, M.~Atzmueller, M.~Kibanov, and G.~Stumme.
\newblock Predictability of evolving contacts and triadic closure in human
  face-to-face proximity networks.
\newblock {\em Social Network Analysis and Mining}, 4(1):217, 2014.
\newblock \href {https://doi.org/10.1007/s13278-014-0217-1}
  {\path{doi:10.1007/s13278-014-0217-1}}.

\bibitem{hartenstein2008tutorial}
H.~Hartenstein and K.~P. Laberteaux.
\newblock A tutorial survey on vehicular ad hoc networks.
\newblock {\em IEEE Communications Magazine}, 46(6):164, 2008.
\newblock \href {https://doi.org/10.1109/MCOM.2008.4539481}
  {\path{doi:10.1109/MCOM.2008.4539481}}.

\bibitem{morales2010building}
J.~M. Morales, P.~R. Moorcroft, J.~Matthiopoulos, J.~L. Frair, J.~G. Kie, R.~A.
  Powell, E.~H. Merrill, and D.~T. Haydon.
\newblock Building the bridge between animal movement and population dynamics.
\newblock {\em Philosophical Transactions of the Royal Society B}, 365(1550):22891, 2010.
\newblock \href {https://doi.org/10.1098/rstb.2010.0082}
  {\path{doi:10.1098/rstb.2010.0082}}.

\bibitem{fink2013robust}
J.~Fink, A.~Ribeiro, and V.~Kumar.
\newblock Robust control of mobility and communications in autonomous robot
  teams.
\newblock {\em IEEE Access}, 1:290, 2013.
\newblock \href {https://doi.org/10.1109/access.2013.2262013}
  {\path{doi:10.1109/access.2013.2262013}}.

\bibitem{de2011world}
L.~De~Benedictis and L.~Tajoli.
\newblock The world trade network.
\newblock {\em The World Economy}, 34(8):1417, 2011.
\newblock \href {https://doi.org/10.1111/j.1467-9701.2011.01360.x}
  {\path{doi:10.1111/j.1467-9701.2011.01360.x}}.

\bibitem{volberda2003co}
H.~W. Volberda and A.~Y. Lewin.
\newblock Co-evolutionary dynamics within and between firms: From evolution to
  co-evolution.
\newblock {\em Journal of Management Studies}, 40(8):2111, 2003.
\newblock \href {https://doi.org/10.1046/j.1467-6486.2003.00414.x}
  {\path{doi:10.1046/j.1467-6486.2003.00414.x}}.

\bibitem{kennedy2023building}
S.~Kennedy, M.~Wish, P.~Smith, J.~Sherrell, W.~Shields, and R.~Gera.
\newblock Building a reliable, dynamic and temporal synthetic model of the
  world trade web.
\newblock In {\em Complex Networks XIII: Proceedings of the 13th Conference on
  Complex Networks, CompleNet 2022}, p. 69. Springer, 2023.
\newblock \href {https://doi.org/10.1007/978-3-031-17658-6_6}
  {\path{doi:10.1007/978-3-031-17658-6_6}}.

\bibitem{carroll2007evolution}
S.~P.~Carroll, A.~P.~Hendry, D.~N.~Reznick, and C.~W. Fox.
\newblock Evolution on ecological time-scales.
\newblock {\em Functional Ecology}, 21(3):387, 2007.
\newblock \href {https://doi.org/10.1111/j.1365-2435.2007.01289.x}
  {\path{doi:10.1111/j.1365-2435.2007.01289.x}}.

\bibitem{held2014adaptive}
T.~Held, A.~Nourmohammad, and M.~L{\"a}ssig.
\newblock Adaptive evolution of molecular phenotypes.
\newblock {\em Journal of Statistical Mechanics},
  2014(9):P09029, 2014.
\newblock \href {https://doi.org/10.1088/1742-5468/2014/09/P09029}
  {\path{doi:10.1088/1742-5468/2014/09/P09029}}.

\bibitem{hendry2017eco}
A.~P. Hendry.
\newblock {\em Eco-evolutionary Dynamics}.
\newblock Princeton University Press, Princeton, 2017.

\bibitem{holme2012temporal}
P.~Holme and J.~Saram{\"a}ki.
\newblock Temporal networks.
\newblock {\em Physics Reports}, 519(3):97, 2012.
\newblock \href {https://doi.org/10.1007/978-3-642-36461-7}
  {\path{doi:10.1007/978-3-642-36461-7}}.

\bibitem{masuda2020guide}
N.~Masuda and R.~Lambiotte.
\newblock {\em A Guide to Temporal Networks}, 2nd edition.
\newblock World Scientific, Singapore, 2020.
\newblock \href {https://doi.org/10.1142/q0268} {\path{doi:10.1142/q0268}}.

\bibitem{bernasconi1987low}
J.~Bernasconi.
\newblock Low autocorrelation binary sequences: statistical mechanics and
  configuration space analysis.
\newblock {\em Journal de Physique}, 48(4):559, 1987.
\newblock \href {https://doi.org/10.1051/jphys:01987004804055900}
  {\path{doi:10.1051/jphys:01987004804055900}}.

\bibitem{jones2012autocorrelation}
S.~D. Jones, C.~Le~Qu{\'e}r{\'e}, and C.~R{\"o}denbeck.
\newblock Autocorrelation characteristics of surface ocean pCO2and air-sea CO2
  fluxes.
\newblock {\em Global Biogeochemical Cycles}, 26(2), 2012.
\newblock \href {https://doi.org/10.1029/2010gb004017}
  {\path{doi:10.1029/2010gb004017}}.

\bibitem{sokal1978spatial}
R.~R. Sokal and N.~L. Oden.
\newblock Spatial autocorrelation in biology: 1. methodology.
\newblock {\em Biological Journal of the Linnean Society}, 10(2):199,
  1978.
\newblock \href {https://doi.org/10.1111/j.1095-8312.1978.tb00013.x}
  {\path{doi:10.1111/j.1095-8312.1978.tb00013.x}}.

\bibitem{porte2014autocorrelation}
X.~Porte, O.~D'Huys, T.~J{\"u}ngling, D.~Brunner, M.~C. Soriano, and
  I.~Fischer.
\newblock Autocorrelation properties of chaotic delay dynamical systems: A
  study on semiconductor lasers.
\newblock {\em Physical Review E}, 90(5):052911, 2014.
\newblock \href {https://doi.org/10.1103/physreve.90.052911}
  {\path{doi:10.1103/physreve.90.052911}}.

\bibitem{agterberg1970autocorrelation}
F.~P. Agterberg.
\newblock Autocorrelation functions in geology.
\newblock In {\em Geostatistics: a colloquium}, p. 113. Springer, 1970.
\newblock \href {https://doi.org/10.1007/978-1-4615-7103-2_10}
  {\path{doi:10.1007/978-1-4615-7103-2_10}}.

\bibitem{wise1955autocorrelation}
J.~Wise.
\newblock The autocorrelation function and the spectral density function.
\newblock {\em Biometrika}, 42(1/2):151, 1955.
\newblock \href {https://doi.org/10.2307/2333432} {\path{doi:10.2307/2333432}}.

\bibitem{clauset2012persistence}
A.~Clauset and N.~Eagle, 
\newblock ``Persistence and periodicity in a dynamic proximity network,'' 
in \textit{Proceedings of the DIMACS/DyDAn Workshop on Computational Methods for Dynamic Interaction Networks}, Piscataway, NJ, 2007.
\newblock URL: \url{https://arxiv.org/abs/1211.7343}.

\bibitem{williams2019effects}
O.~E. Williams, F.~Lillo, and V.~Latora.
\newblock Effects of memory on spreading processes in non-Markovian temporal
  networks.
\newblock {\em New Journal of Physics}, 21(4):043028, 2019.
\newblock \href {https://doi.org/10.1088/1367-2630/ab13fb}
  {\path{doi:10.1088/1367-2630/ab13fb}}.

\bibitem{lacasa2022correlations}
L.~Lacasa, J.~P. Rodriguez, and V.~M. Egu\'{i}luz.
\newblock Correlations of network trajectories.
\newblock {\em Physical Review Research}, 4(4):L042008, 2022.
\newblock \href {https://doi.org/10.1103/PhysRevResearch.4.L042008}
  {\path{doi:10.1103/PhysRevResearch.4.L042008}}.

\bibitem{jo2024temporal}
H.-H. Jo, T.~Birhanu, and N.~Masuda.
\newblock Temporal scaling theory for bursty time series with clusters of
  arbitrarily many events.
\newblock {\em Chaos},
  34(8):083110, 2024.
\newblock \href {https://doi.org/10.1063/5.0219561}
  {\path{doi:10.1063/5.0219561}}.

\bibitem{Nicosia2013}
V.~Nicosia, J.~Tang, C.~Mascolo, M.~Musolesi, G.~Russo, and V.~Latora.
\newblock {\em Graph Metrics for Temporal Networks}, p. 15.
\newblock Springer-Verlag, Berlin, 2013.
\newblock \href {https://doi.org/10.1007/978-3-642-36461-7_2}
  {\path{doi:10.1007/978-3-642-36461-7_2}}.

\bibitem{buttner2016adaption}
K.~B{\"u}ttner, J.~Salau, and J.~Krieter.
\newblock Adaption of the temporal correlation coefficient calculation for
  temporal networks (applied to a real-world pig trade network).
\newblock {\em SpringerPlus}, 5(1):1, 2016.
\newblock \href {https://doi.org/10.1186/s40064-016-1811-7}
  {\path{doi:10.1186/s40064-016-1811-7}}.

\bibitem{navidi2004stationary}
W.~Navidi and T.~Camp.
\newblock Stationary distributions for the random waypoint mobility model.
\newblock {\em IEEE transactions on Mobile Computing}, 3(1):99, 2004.
\newblock \href {https://doi.org/10.1109/tmc.2004.1261820}
  {\path{doi:10.1109/tmc.2004.1261820}}.

\bibitem{sinclair2010mobile}
A.~Sinclair and A.~Stauffer.
\newblock Mobile geometric graphs, and detection and communication problems in
  mobile wireless networks.
\newblock {\em Preprint arXiv:1005.1117}, 2010.
\newblock URL: \url{https://arxiv.org/abs/1005.1117}.

\bibitem{fujiwara2011synchronization}
N.~Fujiwara, J.~Kurths, and A.~D{\'\i}az-Guilera.
\newblock Synchronization in networks of mobile oscillators.
\newblock {\em Physical Review E}, 83(2):025101, 2011.
\newblock \href {https://doi.org/10.1103/physreve.83.025101}
  {\path{doi:10.1103/physreve.83.025101}}.

\bibitem{rhee2011levy}
I.~Rhee, M.~Shin, S.~Hong, K.~Lee, S.~J. Kim, and S.~Chong.
\newblock On the levy-walk nature of human mobility.
\newblock {\em IEEE/ACM Transactions on Networking}, 19(3):630, 2011.
\newblock \href {https://doi.org/10.1109/infocom.2008.145}
  {\path{doi:10.1109/infocom.2008.145}}.

\bibitem{birand2011dynamic}
B.~Birand, M.~Zafer, G.~Zussman, and K.-W. Lee.
\newblock Dynamic graph properties of mobile networks under levy walk mobility.
\newblock In {\em 2011 IEEE Eighth International Conference on Mobile Ad-hoc
  and Sensor Systems}, p.~292. IEEE, 2011.
\newblock \href {https://doi.org/10.1109/MASS.2011.36}
  {\path{doi:10.1109/MASS.2011.36}}.

\bibitem{van1997dynamic}
J.~van~den Berg, R.~Meester, and D.~G. White.
\newblock Dynamic boolean models.
\newblock {\em Stochastic Processes and their Applications}, 69(2):247,
  1997.
\newblock \href {https://doi.org/10.1016/S0304-4149(97)00044-6}
  {\path{doi:10.1016/S0304-4149(97)00044-6}}.

\bibitem{konstantopoulos2014mobile}
T.~Konstantopoulos.
\newblock The mobile boolean model: an overview and further results.
\newblock {\em Preprint arXiv:1407.6834}, 2014.
\newblock URL: \url{https://arxiv.org/abs/1407.6834}.

\bibitem{ramanathan2002brief}
R.~Ramanathan and J.~Redi.
\newblock A brief overview of ad hoc networks: challenges and directions.
\newblock {\em IEEE Communications Magazine}, 40(5):20, 2002.
\newblock \href{https://doi.org/10.1109/MCOM.2002.1006968}
  {\path{doi:10.1109/MCOM.2002.1006968}}.

\bibitem{zhang2014connectivity}
Y.~Zhang, H.~Zhang, W.~Sun, and C.~Pan.
\newblock Connectivity analysis for vehicular ad hoc network based on the
  exponential random geometric graphs.
\newblock In {\em 2014 IEEE Intelligent Vehicles Symposium Proceedings}, p.~993. IEEE, 2014.
\newblock \href{https://doi.org/10.1109/IVS.2014.6856464}
  {\path{doi:10.1109/IVS.2014.6856464}}.

\bibitem{shang2009exponential}
Y.~Shang.
\newblock Exponential random geometric graph process models for mobile wireless
  networks.
\newblock In {\em 2009 International Conference on Cyber-enabled Distributed
  Computing and Knowledge Discovery}, p. 56. IEEE, 2009.
\newblock \href{https://doi.org/10.1109/CYBERC.2009.5342212}
  {\path{doi:10.1109/CYBERC.2009.5342212}}.

\bibitem{grimmett2022brownian}
G.~R. Grimmett and Z.~Li.
\newblock Brownian snails with removal: epidemics in diffusing populations.
\newblock {\em Electronic Journal of Probability}, 27:1, 2022.
\newblock \href{https://doi.org/10.1214/22-EJP804}{\path{doi:10.1214/22-EJP804}}.

\bibitem{clementi2015parsimonious}
A.~Clementi and R.~Silvestri.
\newblock Parsimonious flooding in geometric random-walks.
\newblock {\em Journal of Computer and System Sciences}, 81(1):219, 2015.
\newblock \href{https://doi.org/10.1016/j.jcss.2014.06.002}{\path{doi:10.1016/j.jcss.2014.06.002}}.

\bibitem{zhang2013gossip}
H.~Zhang, Z.~Zhang, and H.~Dai.
\newblock Gossip-based information spreading in mobile networks.
\newblock {\em IEEE Transactions on Wireless Communications},
  12(11):5918, 2013.
\newblock \href {https://doi.org/10.1109/twc.2013.100113.130619}
  {\path{doi:10.1109/twc.2013.100113.130619}}.

\bibitem{ferri2023three}
I.~Ferri, A.~Gaya-{\`A}vila, and A.~D{\'\i}az-Guilera.
\newblock Three-state opinion model with mobile agents.
\newblock {\em Chaos},
  33(9):093121, 2023.
\newblock \href {https://doi.org/10.1063/5.0152674}
  {\path{doi:10.1063/5.0152674}}.

\bibitem{mauve2001survey}
M.~Mauve, J.~Widmer, and H.~Hartenstein, ``A survey on position-based routing in
  mobile ad hoc networks,'' \emph{IEEE Network}, 15(6):30, 2001.
\newblock \href {https://doi.org/10.1109/65.967595}{\path{doi:10.1109/65.967595}}.
  
\bibitem{ye2003framework}
Z.~Ye, S.~V. Krishnamurthy, and S.~K. Tripathi, ``A framework for reliable
  routing in mobile ad hoc networks,'' in \emph{IEEE INFOCOM 2003. Twenty-second
  Annual Joint Conference of the IEEE Computer and Communications Societies
  (IEEE Cat. No. 03CH37428)}, vol.~1, 2003, p. 270.
\newblock \href {https://doi.org/10.1109/INFCOM.2003.1208679}{\path{doi:10.1109/INFCOM.2003.1208679}}.

\bibitem{gu2018ability}
C.~Gu, I.~Downes, O.~Gnawali, and L.~Guibas.
\newblock On the ability of mobile sensor networks to diffuse information.
\newblock In {\em 2018 17th ACM/IEEE International Conference on Information
  Processing in Sensor Networks (IPSN)}, p.~37. IEEE, 2018.
\newblock \href {https://doi.org/10.1109/IPSN.2018.00011}
  {\path{doi:10.1109/IPSN.2018.00011}}.

\bibitem{cheliotis2020simple}
D~Cheliotis, I.~Kontoyiannis, M.~Loulakis, and S.~Toumpis.
\newblock A simple network of nodes moving on the circle.
\newblock {\em Random Structures \& Algorithms}, 57(2):317, 2020.
\newblock \href {https://doi.org/10.1002/rsa.20932}
  {\path{doi:10.1002/rsa.20932}}.

\bibitem{dall2002random}
J.~Dall and M.~Christensen.
\newblock Random geometric graphs.
\newblock {\em Physical Review E}, 66(1):016121, 2002.
\newblock \href {https://doi.org/10.1103/physreve.66.016121}
  {\path{doi:10.1103/physreve.66.016121}}.

\bibitem{hethcote1978immunization}
H.~W. Hethcote.
\newblock An immunization model for a heterogeneous population.
\newblock {\em Theoretical Population Biology}, 14(3):338, 1978.
\newblock \href {https://doi.org/10.1016/0040-5809(78)90011-4}
  {\path{doi:10.1016/0040-5809(78)90011-4}}.

\bibitem{may1984spatial}
R.~M. May and R.~M. Anderson.
\newblock Spatial heterogeneity and the design of immunization programs.
\newblock {\em Mathematical Biosciences}, 72(1):83, 1984.
\newblock \href {https://doi.org/10.1016/0025-5564(84)90063-4}
  {\path{doi:10.1016/0025-5564(84)90063-4}}.

\bibitem{hanski1998metapopulation}
I.~Hanski.
\newblock Metapopulation dynamics.
\newblock {\em Nature}, 396(6706):41, 1998.
\newblock \href {https://doi.org/10.1038/23876} {\path{doi:10.1038/23876}}.

\bibitem{colizza2007reaction}
V.~Colizza, R.~Pastor-Satorras, and A.~Vespignani.
\newblock Reaction--diffusion processes and metapopulation models in
  heterogeneous networks.
\newblock {\em Nature Physics}, 3(4):276, 2007.
\newblock \href {https://doi.org/10.1038/nphys560}
  {\path{doi:10.1038/nphys560}}.

\bibitem{masuda2017random}
N.~Masuda, M.~A. Porter, and R.~Lambiotte.
\newblock Random walks and diffusion on networks.
\newblock {\em Physics Reports}, 716:1, 2017.
\newblock \href {https://doi.org/10.1016/j.physrep.2017.07.007}
  {\path{doi:10.1016/j.physrep.2017.07.007}}.

\bibitem{masuda2005geographical}
N.~Masuda, H.~Miwa, and N.~Konno.
\newblock {Geographical threshold graphs with small-world and scale-free
  properties}.
\newblock {\em Physical Review E}, 71(3):036108, 2005.
\newblock \href {https://doi.org/10.1103/physreve.71.036108}
  {\path{doi:10.1103/physreve.71.036108}}.

\bibitem{hagberg2006designing}
A. Hagberg, P.~J. Swart, and D.~A. Schult.
\newblock Designing threshold networks with given structural and dynamical
  properties.
\newblock {\em Physical Review E}, 74(5):056116, 2006.
\newblock \href {https://doi.org/10.1103/PhysRevE.74.056116}
  {\path{doi:10.1103/PhysRevE.74.056116}}.

\bibitem{campeau2022evolutionary}
W.~Campeau, A.~M. Simons, and B.~Stevens.
\newblock The evolutionary maintenance of L{\'e}vy flight foraging.
\newblock {\em PLoS Computational Biology}, 18(1):e1009490, 2022.
\newblock \href {https://doi.org/10.1371/journal.pcbi.1009490}
  {\path{doi:10.1371/journal.pcbi.1009490}}.

\bibitem{weeks1995observation}
E.~R. Weeks, T.~H. Solomon, J.~S. Urbach, and H.~L. Swinney, \newblock Observation of anomalous diffusion and L{\'e}vy flights, in \emph{L{\'e}vy Flights and Related Topics in Physics}, M.~F. Shlesinger, G.~M. Zaslavsky, and U.~Frisch, Eds., Springer, Berlin, 1995, p. 51.
\newblock \href {https://doi.org/10.1007/3-540-59222-9_25}
  {\path{doi:10.1007/3-540-59222-9_25}}.

\bibitem{feller1991}
W.~Feller, \emph{An Introduction to Probability Theory and Its Applications, Volume 1}, 3rd~ed.
\newblock John Wiley, New York, 1991.

\bibitem{ben2000diffusion}
D. Ben-Avraham and S. Havlin.
\newblock {\em Diffusion and Reactions in Fractals and Disordered Systems}.
\newblock Cambridge University Press, Cambridge, 2000.
\newblock \href {https://doi.org/10.1017/CBO9780511605826}
  {\path{doi:10.1017/CBO9780511605826}}.

\bibitem{kyprianou2014fluctuations}
A.~E. Kyprianou, 
\newblock {\em Fluctuations of L{\'e}vy Processes with Applications: Introductory Lectures}, 2nd~ed.
\newblock Springer-Verlag, Berlin, 2014.
\newblock \href{https://doi.org/10.1007/978-3-642-37632-0}{\path{doi:10.1007/978-3-642-37632-0}}.

\bibitem{barabasi1999emergence}
A.-L. Barab{\'a}si and R.~Albert.
\newblock Emergence of scaling in random networks.
\newblock {\em Science}, 286(5439):509, 1999.
\newblock \href {https://doi.org/10.1126/science.286.5439.509}
  {\path{doi:10.1126/science.286.5439.509}}.

\bibitem{caligiuri2023lyapunov}
A.~Caligiuri, V.~M. Egu{\'\i}luz, L.~Di~Gaetano, T.~Galla, and L.~Lacasa.
\newblock Lyapunov exponents for temporal networks.
\newblock {\em Physical Review E}, 107(4):044305, 2023.
\newblock \href {https://doi.org/10.1103/PhysRevE.107.044305}
  {\path{doi:10.1103/PhysRevE.107.044305}}.

\bibitem{kim2018review}
B.~Kim, K.~H. Lee, L.~Xue, and X.~Niu.
\newblock A review of dynamic network models with latent variables.
\newblock {\em Statistics Surveys}, 12:105, 2018.
\newblock \href {https://doi.org/10.1214/18-SS121}
  {\path{doi:10.1214/18-SS121}}.

\bibitem{perra2012activity}
N.~Perra, B.~Gon{\c{c}}alves, R.~Pastor-Satorras, and A.~Vespignani.
\newblock Activity driven modeling of time varying networks.
\newblock {\em Scientific Reports}, 2(1):469, 2012.
\newblock \href {https://doi.org/10.1038/srep00469}
  {\path{doi:10.1038/srep00469}}.

\bibitem{zino2018modeling}
L.~Zino, A.~Rizzo, and M.~Porfiri.
\newblock Modeling memory effects in activity-driven networks.
\newblock {\em SIAM Journal on Applied Dynamical Systems}, 17(4):2830,
  2018.
\newblock \href {https://doi.org/10.1137/18m1171485}
  {\path{doi:10.1137/18m1171485}}.

\bibitem{sheng2023constructing}
A.~Sheng, Q.~Su, A.~Li, L.~Wang, and J.~B. Plotkin.
\newblock Constructing temporal networks with bursty activity patterns.
\newblock {\em Nature Communications}, 14(1):7311, 2023.
\newblock \href {https://doi.org/10.1038/s41467-023-42868-1}
  {\path{doi:10.1038/s41467-023-42868-1}}.

\bibitem{ye2010optimal}
Z.~Ye and A.~A. Abouzeid.
\newblock Optimal stochastic location updates in mobile ad hoc networks.
\newblock {\em IEEE Transactions on Mobile Computing}, 10(5):638, 2010.
\newblock \href {https://doi.org/10.1109/tmc.2010.201}
  {\path{doi:10.1109/tmc.2010.201}}.

\bibitem{roy2011handbook}
R.~R. Roy, \emph{Handbook of Mobile Ad Hoc Networks for Mobility Models}, 1st~ed. 
\newblock Springer, New York, 2011.
\newblock\href{https://doi.org/10.1007/978-1-4419-6050-4}{\path{doi:10.1007/978-1-4419-6050-4}}.

\bibitem{barbosa2018human}
H.~Barbosa, M.~Barthelemy, G.~Ghoshal, C.~R. James, M.~Lenormand, T.~Louail,
  R.~Menezes, J.~J. Ramasco, F.~Simini, and M.~Tomasini.
\newblock Human mobility: Models and applications.
\newblock {\em Physics Reports}, 734:1, 2018.
\newblock \href {https://doi.org/10.1016/j.physrep.2018.01.001}
  {\path{doi:10.1016/j.physrep.2018.01.001}}.

\bibitem{alessandretti2020scales}
L.~Alessandretti, U.~Aslak, and S.~Lehmann.
\newblock The scales of human mobility.
\newblock {\em Nature}, 587(7834):402, 2020.
\newblock \href {https://doi.org/10.1038/s41586-020-2909-1}
  {\path{doi:10.1038/s41586-020-2909-1}}.

\bibitem{edsberg2022understanding}
P.~E.~M{\o}llgaard, S.~Lehmann, and L.~Alessandretti.
\newblock Understanding components of mobility during the COVID-19 pandemic.
\newblock {\em Philosophical Transactions of the Royal Society A},
  380(2214):20210118, 2022.
\newblock \href {https://doi.org/10.1098/rsta.2021.0118}
  {\path{doi:10.1098/rsta.2021.0118}}.

\bibitem{bettstetter2004stochastic}
C.~Bettstetter, H.~Hartenstein, and X.~P{\'e}rez-Costa.
\newblock Stochastic properties of the random waypoint mobility model.
\newblock {\em Wireless Networks}, 10(5):555, 2004.
\newblock \href {https://doi.org/10.1023/b:wine.0000036458.88990.e5}
  {\path{doi:10.1023/b:wine.0000036458.88990.e5}}.

\bibitem{hyytia2006spatial}
E.~Hyytia, P.~Lassila, and J.~Virtamo.
\newblock Spatial node distribution of the random waypoint mobility model with
  applications.
\newblock {\em IEEE Transactions on Mobile Computing}, 5(6):680, 2006.
\newblock \href {https://doi.org/10.1109/tmc.2006.86}
  {\path{doi:10.1109/tmc.2006.86}}.

\bibitem{vanderkamp1977gravity}
J.~Vanderkamp.
\newblock The gravity model and migration behaviour: An economic
  interpretation.
\newblock {\em Journal of Economic Studies}, 4(2):89, 1977.
\newblock \href {https://doi.org/10.1108/eb002472}
  {\path{doi:10.1108/eb002472}}.

\bibitem{beine2016practitioners}
M.~Beine, S.~Bertoli, and J.~Fern{\'a}ndez-Huertas~Moraga.
\newblock A practitioners' guide to gravity models of international
  migration.
\newblock {\em The World Economy}, 39(4):496, 2016.
\newblock \href {https://doi.org/10.1111/twec.12265}
  {\path{doi:10.1111/twec.12265}}.

\bibitem{camp2002survey}
T.~Camp, J.~Boleng, and V.~Davies.
\newblock A survey of mobility models for ad hoc network research.
\newblock {\em Wireless Communications and Mobile Computing}, 2(5):483,
  2002.
\newblock \href {https://doi.org/10.1002/wcm.72} {\path{doi:10.1002/wcm.72}}.

\bibitem{feller1991vol2}
W.~Feller, \emph{An Introduction to Probability Theory and Its Applications, Volume 2}, 2nd~ed.
\newblock John Wiley, New York, 1991.

\bibitem{okada2020long}
M.~Okada, K.~Yamanishi, and N.~Masuda.
\newblock Long-tailed distributions of inter-event times as mixtures of
  exponential distributions.
\newblock {\em Royal Society Open Science}, 7(2):191643, 2020.
\newblock \href {https://doi.org/10.1098/rsos.191643}
  {\path{doi:10.1098/rsos.191643}}.

  \bibitem{jiang2016two}
Z.~Q. Jiang, W.~J. Xie, M.~X. Li, W.~X. Zhou, and D.~Sornette, Two-state
  Markov-chain Poisson nature of individual cellphone call statistics,
  \emph{Journal of Statistical Mechanics}, 2016(7):073210, 2016. \newblock \href{https://doi.org/10.1088/1742-5468/2016/07/073210}
  {\path{doi:10.1088/1742-5468/2016/07/073210}}.


\bibitem{feldmann1998fitting}
A.~Feldmann and W.~Whitt, Fitting mixtures of exponentials to long-tail
  distributions to analyze network performance models, \emph{Performance
  Evaluation}, 31(3--4):245, 1998. \newblock \href{https://doi.org/10.1016/S0166-5316(97)00003-5}
  {\path{doi:10.1016/S0166-5316(97)00003-5}}.



\end{thebibliography}
\end{document}